\documentclass[aip]{revtex4-1}
\usepackage{graphicx}
\usepackage{color}
\usepackage[utf8]{inputenc}
\usepackage{caption}
\def\@email#1#2{%
 \endgroup
 \patchcmd{\titleblock@produce}
  {\frontmatter@RRAPformat}
  {\frontmatter@RRAPformat{\produce@RRAP{*#1\href{mailto:#2}{#2}}}\frontmatter@RRAPformat}
  {}{}
}%

\draft % marks overfull lines with a black rule on the right

\begin{document}

%\preprint{}

\title{Viscous effects on nonlinear double tearing mode and plasmoid formation in adjacent Harris sheets} %Title of paper

%\author{Nisar AHMAD$^{1,2}$, Ping ZHU$^{3,4\ast}$, Chao Shen$^{1\ast}$,Ahmad ALI$^5$, and Shiyong ZENG$^6$}
\author{Nisar AHMAD}
 \altaffiliation{Formerly at University of Science and Technology of China}
\affiliation{School of Science, Harbin Institute of Technology (Shenzhen), Shenzhen 518055, People’s Republic of China}
%\address{\small CAS Key Laboratory of Geospace Environment and Department of Engineering and Applied Physics, University of Science and Technology of China, Hefei 230026, People's Republic of China}
\author{Ping ZHU}
 \altaffiliation{zhup@hust.edu.cn and shenchao@hit.edu.cn} 
\affiliation{ International Joint Research Laboratory of Magnetic Confinement Fusion and Plasma Physics, State Key Laboratory of Advanced Electromagnetic Engineering and Technology, School of Electrical and Electronic Engineering, Huazhong University of Science and Technology, Wuhan, Hubei 430074, People's Republic of China}
\affiliation{Department of Engineering Physics, University of Wisconsin-Madison, Madison, Wisconsin 53706, USA}
\author{Chao SHEN$^{\rm{b)}}$}
%\email{shenchao@hit.edu.cn}
\affiliation{College of Science, Harbin Institute of Technology (Shenzhen), Shenzhen 518055, People’s Republic of China}
\author{Ahmad ALI}
\affiliation{Pakistan Tokamak Plasma Research Institute, Islamabad 3329, Pakistan}
\author{Shiyong ZENG}
\affiliation{ Department of Engineering and Applied Physics, University of Science and Technology of China, Hefei, Anhui 230026, People's Republic of China}
%\ead{zhup@hust.edu.cn}
\vspace{10pt}
%{zhup@hust.edu.cn}{shenchao@hit.edu.cn}
%\author{}
%\thanks{}
%\affiliation{}

% Collaboration name, if desired (requires use of superscript address option in \documentclass). 

\date{\today}

\begin{abstract}
In this paper, we study the effects of viscosity on the evolution of double tearing mode (DTM) in a pair of adjacent Harris sheets based on the resistive MHD model in the NIMROD code. Similar to the tearing mode in the conventional single Harris sheet, a transition is observed in the generation of both normal and monster plasmoids at ${P_r = 1}$. In the ${P_r < 1}$ regime of DTM, normal plasmoids (small plasmoids) are generated along with monster plasmoid, whereas in the single tearing mode (STM) cases such a generation is not observed. When ${P_r}$ is above the critical value, the generation of monster plasmoid is halted. Correspondingly, in the ${P_r < 1}$ regime, a quadrupolar flow advects along poloidal direction, but in ${P_r > 1}$ regime this flow advection is inhibited.
\end{abstract}


\maketitle %\maketitle must follow title, authors, abstract and \pacs

\section{Introduction}
%================
% paragraph 1
%================
In space and fusion plasmas \cite{bierwage2005fast,wang2011interlocking,ishii2009plasma,akramov2017non}, double tearing modes (DTMs) and associated fast magnetic reconnection mechanisms have been subjects of broad interests in recent years. Double (multiple) current sheets are commonly observed in space plasma, such as in neighboring coronal helmet streamers \cite{dahlburg1995triple}, the solar wind \cite{crooker1993multiple}, and the solar corona \cite{mikic1988dynamical} in general. The Earth's bow shock \cite{schwartz1988active}, for example, can trigger double, triple, or even multiple tearing modes \cite{yan1994tearing,bowers2007spectral,baty2017explosive}. Reversed magnetic shear (RMS) are commonly formed in advanced scenarios of tokamak operations \cite{levinton1995improved}, which can lead to the onset of double tearing modes on two adjacent rational surfaces with the same safety factor and the subsequent off-axis sawtooth oscillation or even disruption \cite{ishii2002structure,wang2007fast,janvier2011structure,priest1985magnetohydrodynamics,chang1996off}. The magnetic reconfiguration caused by DTM has been extensively studied \cite{ishii2002structure,ishii2003long,wang2007fast,wang2008shear}, which is among the most significant aspects of DTM's nonlinear evolution.

The linear growth rate of DTM for small distances between adjecent current sheets scales with the resistivity ${\eta}$ as ${\gamma_{lin} \sim \eta^{1/3}}$. When the distance between them is larger, scaling becomes ${\gamma_{lin} \sim \eta^{3/5}}$ \cite{pritchett1980linear}. In the nonlinear regime, the coupling in DTM would disrupt linearly stable tearing modes and ramp up the growth of magnetic islands \cite{ishii2002structure,yu1996nonlinear,persson1994nonlinear}. Effects like anomalous electron viscosity, finite beta, and sheared toroidal flows can make DTMs more unstable \cite{furth1973tearing,dong2003double,held1999magnetohydrodynamic}.

Many experts have studied the nonlinear effects of DTMs due to resistivity, shear flow, large guiding field, and current sheet spacing \cite{ma2017effect,zhang2011nonlinear,nemati2018unstable,shen1998magnetic,shen1998properties}. During the nonlinear evolution of DTM, unstable long and thin current sheets can form as a result of the island's interaction and breakup, leading to secondary islands \cite{zhang2011nonlinear} or even tertiary islands \cite{nemati2016formation,nemati2017dynamics}, demonstrating the variety and complexity of physics in the high Lundquist number regime.

In recent studies \cite{nemati2016formation,Guo2017}, multiple plasmoid formations were observed in the absence of viscosity in a simulation of multiple current sheet systems with high Lundquist number parameters. However, there has been limited study about viscosity effects on DTM nonlinear evolution. Minor electromagnetic perturbations may generate anomalous electron viscosity in tokamak plasmas \cite{konovalov2002transport}. As a kind of dissipation effect, anomalous electron viscosity may also drive tearing mode instability, which is thought to be one of the possible causes of plasma disruption \cite{kaw1979tearing,ahmad2021viscous,ALI2015,ALI2019}. The electron viscosity tearing modes are one potential physical mechanism for the rapid saw-tooth disruption reported in experiments \cite{aydemir1990magnetohydrodynamic}. In slab configuration plasmas, the linear features of DTMs owing to anomalous electron viscosity have also been stated \cite{dong2003double}.

In this work, the resistive-viscous MHD model implemented in the NIMROD code \cite{sovinec2004nonlinear} is utilized to systematically investigate the nonlinear evolution of DTM, focusing on the effect of viscosity on the formation and evolution of plasmoids. Normal plasmoids (small plasmoids) \cite{nemati2017dynamics} along with the monster plasmoid appear in the ${P_r < 1}$ regime, whereas in the ${P_r > 1}$ regime only monster plasmoid appears. The plasmoid dynamics in our simulations are compared with previous studies \cite{He2015, Mao2021}. 

The rest of the paper is organized as follows. In Sec. II, we briefly describe our simulation model. In Sec. III and IV, both linear and nonlinear simulation results are reported respectively. At the end in Sec. V, summary and
discussion are presented.

\section{Simulation model and equilibrium} 

Our simulations are based on the single-fluid full MHD model implemented in the NIMROD \cite{sovinec2004nonlinear} code. The below visco-resistive MHD model (1)–(4) equations used in our study are all non-dimensional.

\begin{equation}
  \label{eq=one}
\frac{\partial \rho}{\partial t}+\nabla \cdot(\rho \mathbf{v})=0
\end{equation}

\begin{equation}
  \label{eq=one}
\rho\left(\frac{\partial}{\partial t}+\mathbf{v} \cdot \nabla\right) \mathbf{v}=\mathbf{J} \times \mathbf{B}-\nabla p+\rho \nu \nabla^{2} \mathbf{v}
\end{equation}

\begin{equation}
  \label{eq=one}
\frac{N}{\gamma-1}\left(\frac{\partial}{\partial t}+\mathbf{v} \cdot \nabla\right) \mathbf{T}=-{\frac{p}{2}} \nabla \cdot \mathbf{v}
\end{equation}

\begin{equation}
  \label{eq=one}
\frac{\partial \mathbf{B}}{\partial t}=\nabla \times(\mathbf{v} \times \mathbf{B}-\eta\mathbf{J})=\nabla \times(\mathbf{v} \times \mathbf{B}-\eta\nabla \times \mathbf{B})
\end{equation}
where ${\rho,\mathbf{J},p,\gamma,\mathbf{v},N,\mathbf{B},\nu}$ and ${\eta}$ are the plasma mass density, current density, pressure, specific heat ratio, velocity, number density, magnetic field, viscosity, and resistivity, respectively. The Boltzmann constant (k) has been absorbed into temperature. The particle mass density ${\rho}$ and number density ${N}$ are related through the mass per ion, and total temperature and pressure relate the ideal gas relation, ${p}$ = 2${NT}$, considering rapid thermal equilibration and quasi-neutrality condition among electrons and ions. The initial equilibrium profiles for ${B_{0z}(x)}$ , ${J_{0y}(x)}$ and pressure are given below.

\begin{equation}
  \label{eq=one}
{{B}}_{0 z}\left({x}\right)= 1-\tanh \left(\frac{x+x_0}{a}\right)+\tanh \left(\frac{x-x_0}{a}\right)
\end{equation}

\begin{equation}
  \label{eq=one}
{{J}}_{0 y}\left({x}\right)= \frac{1}{a} {\mathrm{sech^2}} \left(\frac{x-x_0}{a}\right)-\frac{1}{a} {\mathrm{sech^2}} \left(\frac{x+x_0}{a}\right)
\end{equation}

\begin{equation}
  \label{eq=one}
{p}_{0}\left({x}\right)= T_0 + 0.5 \left(1-\tanh \left(\frac{x+x_0}{a}\right)+\tanh \left(\frac{x-x_0}{a}\right)\right)^2
\end{equation}
where the Cartesian coordinates ${(x, y, z)}$ are adopted.
The equilibrium profiles of ${B_{0 z}(x)}$ , ${J_{0y}(x)}$ and ${p_{0}(x)}$ at different resonance surfaces are plotted in Fig. 1. 
Here ${a}$ controls the width of the profile, ${x_0}$ represents the half distance between the two adjacent current sheets, and we assume a uniform mass density ${\rho}$. The magnetic field is normalized by the field magnitude ${B_0}$ at the edge of the Harris sheet, i.e. ${x \rightarrow\pm\infty}$. The spatial normalization unit ${x_0 = d/2}$ is the half distance between the two adjacent current sheets. The mass density is normalized by the field magnitude ${\rho_0}$ at the center of the current sheets. The normalization units for time, pressure, and velocity are ${t_0 = \tau_{\mathrm{A0}} = x_0/u_0 = x_0/u_{\mathrm{A0}}}$, ${p_0 = B_0^2/\mu_0}$, and ${u_0 = u_{\mathrm{A0}} = B_0/\sqrt{\mu_0\rho_0}}$ respectively. Also the adiabatic index ${\gamma = 5/3}$ and the heat flux is zero. In the above non-dimensional equations (1)–(4), both the viscosity ${\nu}$ and the resistivity ${\eta}$ are the normalized dimensionless parameters, with ${\nu = \frac{P_\mathrm{r}}{S}}$, and ${\eta = S^{-1}}$, where ${S}$ is the Lundquist number ${S = \tau_\mathrm{R}/\tau_{\mathrm{A0}}}$, ${\tau_\mathrm{R} = \mu_0x_0^2/\eta_\mathrm{D}}$, ${\eta_\mathrm{D}}$ is the unnormalized dimensionless resistivity, and ${P_\mathrm{r} = \mu_0\nu_\mathrm{D}/\eta_\mathrm{D}}$ is the magnetic Prandtl number, ${\nu_\mathrm{D}}$ is the unnormalized dimensionless viscosity, with ${\nu = \nu_\mathrm{D}\tau_{\mathrm{A0}}/x_0^2}$. 

 The resistive MHD equations are solved in a rectangular domain of the form like ${[-L_x, L_x]\times{[-L_y, L_y]}}$. Note that the wave number is defined as ${L_y = 2\pi m/k}$ with ${k}$ is the mode wave number along ${y}$, and ${m}$ being the mode number. Periodic boundary conditions are imposed along the ${y}$ direction and all perturbations in ${x}$ direction are taken to vanish at the boundaries.

 \section{Linear double tearing mode (DTM)}

In Fig. 2, we have plotted contours of the linear mode structure for different fields at ${d = 1}$ between the adjacent current sheets. We have performed these study just to make an overview and conclusion about a specific case to study the nonlinear dynamics for double tearing mode. As we know the reconnection greatly depends on quadrupolar flow cells. At large distances between the rational surfaces these quadrupolar flow cells partially advected in the poloidal direction. To analyze the nonlinear dynamics of double tearing mode, it is necessary to completely learn about these flow cells and the role of these cells in the generation of plasmoids. As a quadrupolar flow can be generated and/or destroyed outside the current sheet. Also this quadrupolar flow directly affected by the viscosity which surely influences the reconnection process. 

In Fig. 3(a), growth rate is plotted as a function of ${\eta}$ for different distances between the adjacent current sheets. For a fixed distance between the current sheets, a good agreement with existing theory is found at small value of ${\eta}$. But for large value of ${\eta}$, it is observed that our results deviate from the existing theory \cite{coppi1976resistive}. From Fig. 3(a), we can also conclude that for a highly unstable system the power of ${\eta}$ remains near to 0.33 which matches the scaling law for small distances between adjacent current sheets \cite{pritchett1980linear}. In Fig. 3(b), for different ${d}$ within the range [0.5, 3], the parameter ${\alpha}$ ( where ${\alpha}$ is $\gamma_{lin}$ = ${\eta^{\alpha}}$) is plotted as a function of the distances between the rational surfaces. The transition between the scaling ${\alpha = 3/5}$ for large ${d}$ and ${\alpha = 1/3}$ for small ${d}$ is clearly observed in previous studies \cite{Shu2013}. In our case, we have selected the anti-symmetric equilibrium, the change in the ${\alpha}$ index is not matching as found in literature. As we increase the distance between the current sheets, the ${\alpha}$ increases up to the distance ${d = 1.6}$, then it decreases quickly. This trend is different from the existing previous results. This maybe due to the fact that we are dealing with highly unstable system.

It is found [Fig. 4(a)], as the viscosity increases, the growth rate quickly slows down. The magnetic Prandtl number ${P_r = \nu / \eta}$ for typical fusion systems can be the order of 100. Also for a fixed value of distance between those rational surfaces, the growth rate shows a transition at ${P_r = 1}$. In ${P_r < 1}$ regime, the effect of viscosity is negligible whereas in ${P_r > 1}$ regime, the growth rate decreases sharply. The viscous scaling for our simulations at ${x_0 = 0.25, 0.5}$ and ${1.25}$ are ${\gamma\sim\nu^{-0.27}}$, ${\gamma\sim\nu^{-0.22}}$ and ${\gamma\sim\nu^{-0.25}}$ respectively. These scaling are very close to the theoretical scaling \cite{ofman1992double,dong2003double}. From Fig. 4(b), for a fixed Prandtl number, the growth rate decreases by increasing the distances between the current sheets or vice versa.  

\section{Nonlinear DTM and plasmoids}
\subsection{Nonlinear DTM evolution}
To study the effects on the nonlinear DTM evolution, one particular case having wave vector ${ka = 0.15}$ is selected. The viscosity is varied in the form of the Prandtl number in the range ${P_r = 0.33,0.5,1,5, 10, 100}$ with the fixed resistivity ${\eta = 0.00028}$ and the current sheets spacing ${d = 1.1}$. ${96 \times 96}$ 2D ﬁnite elements with a 5th order polynomial basis function in each 2D direction are used in our simulations to ensure numerical convergence [Fig. 5].

The time histories of the plasma kinetic
energy for different values of Prandtl number are compared [Fig. 6], where the black arrows indicate the moments of monster plasmoids generation and the red arrows show the subsequent moments of the normal (small) plasmoid formation. In the case of ${P_r = 0.33}$, the kinetic energy grows very fast in the linear
phase and reaches maximum very quickly due to the generation of normal plasmoids [Fig. 6(a)]. By increasing the
Prandtl number, such as ${P_r = 10}$ , the evolution of
kinetic energy goes through four stages: an early linear growth
phase, slow nonlinear Rutherford phase, fast flux-driven
reconnection phase, and a decay phase. In results, by increasing the Prandtl number the early linear growth phase and the slow nonlinear Rutherford phase evolve for longer time. Also by increasing the ${P_r = 100}$, fast flux-driven reconnection phase becomes absent [Fig. 6(c)].

\subsection{Plasmoid formation}

The reconnection rate and plasmoids generation in DTM is different as compared to the STM due to the different mechanisms involved. Comparison between ${P_r = 0.33, 1, 10}$ and ${100}$ shows that the reconnection current sheet is much thinner and longer for ${P_r = 0.33}$ [Fig. 7(a)]. Such a large aspect ratio current sheet is unstable to plasmoid creating tearing modes so that a monster plasmoid produces and grows in size in the middle of the current sheet [Fig. 7(b)]. The monster plasmoid constitutes a secondary instability that divides the primary current ribbon into two adjacent ones on either side of it. As the primary DTM island generated from the left current layer keeps moving, the pairs of subsidiary current sheets are squeezed greatly to become so thin that they also become unstable, each generating plasmoids, as shown in
Fig. 7(c). During their growth, as shown in Fig. 7, all
plasmoids in each current sheet coalesce with each other to form a monster plasmoid at the center of the current sheet [Fig. 7(d)]. This monster plasmoid becomes large enough and moves left, however it can curve and push the primary DTM island significantly. Hence it is inferred
that such a secondary tearing instability on the DTM
reconnection current ribbon leads to the fast growth of kinetic
energy during t = 85 to 100 (Fig. 6). 

But as we increase the Prandtl number further, the possibility for the production of normal plasmoids decreases quickly because of the wider secondary current sheet. Figs. 8 and 9 show the 2D contours of current density and 2D magnetic field lines for the ${P_r = 1}$ and the ${P_r = 10}$ cases respectively. Due to magnetic reconnection, a thin and curving current sheet forms in the island [Figs. 8(a) and 9(a)]. In these cases we only observe the monster plasmoid, and the production of normal plasmoids was not found in these cases. Figs. 8 and 9 show that as the monster plasmoid becomes large enough and moves in the middle, it can curve and push the primary
DTM island significantly. This current sheet finally disappears and the islands positions exchange as shown in Figs. 7(d), 8(d) and 9(d). It is also observed from Fig. 10 that for ${\nu = 0.028}$ i.e. ${P_r = 100}$ case there is no monster plasmoid formation (Fig. 10). This process shows the dissipation and damping nature of large viscosity. Finally the reconnection finishes and we get the straight field lines after the exhaustion of initial flux.

Fig. 11 shows the evolution of monster plasmoid for different ${P_r}$ cases and also summarizes the relationships among monster plasmoid width and ${P_r}$ numbers. At lower ${P_r}$, as we increase the viscosity, the width of monster plasmoid increases sharply up to ${P_r = 1}$. The ${P_r = 1}$ number also separates two regimes for the monster plasmoid width. In the ${P_r < 1}$ regime, the width of plasmoid increases drastically with viscosity, whereas in the ${P_r > 1}$ regime, the width of monster plasmoid decreases sharply with the viscosity. This behaviour of plasmoid dynamics with viscosity is interestingly matching with the linear and nonlinear reconnection dependence on viscosity in STM \cite{ali2014abrupt}.

\subsection{Flow pattern and vortices}

%\textcolor[rgb]{0,0,1}{}.
The flow pattern and vortex structure developed around the magnetic islands and plasmoids show unique features of DTM. The viscous effects on the flow pattern associated with the DTM and the corresponding plasmoid formation are also significant (Figs. 12-15).

In the early stages before the current sheet formation, the magnetic island width are so small that the plasma flows are limited inside of the separated islands. As the island width grows to a certain size, the plasma flows in the islands at the both current sheets merge together to form global vortices [Fig. 12(a)]. After the collapse of current sheet, thin and intensive shear flow layers form and are localized around the plasmoid on each resonant surface. The vortex structures and thin poloidal shear flow layers remain long after the merging of magnetic separatrices [Figs. 12(d) and 12(h)]. As we increase the viscosity up to 0.00028 (${P_r = 1}$), the vortices formation becomes slower and larger in size. At larger viscosity, the reconnection and the production of intensive thin poloidal shear flow layers are more inhibited. This fact may contribute as a mechanism to the oscillating decay of kinetic energy at a considerably faster pace when ${P_r = 10}$ and ${P_r = 100}$ than when ${P_r = 0.33}$ and ${P_r = 1}$.

\section{Summary}

In this paper, we study the viscous effects on the dynamics of the DTM induced plasmoid formation using MHD simulations. The monster plasmoid first appears in the middle of the DTM current sheet when the kinetic energy reaches its ﬁrst peak. Then, normal
plasmoids appear for a small value of the ${P_r}$, making the kinetic energy abruptly increase to a higher level. A transition is observed at ${P_r =1 }$ for the generation of normal plasmoids, which is similar to the STM case. By increasing the distance between the rational surfaces or by decreasing the Prandtl number, the production of normal plasmoids increases or vice versa. Up to a certain critical value of ${P_r}$, the production of monster plasmoid also halts. The overall magnetic and flow structures of DTM as well as the induced formation of plasmoids, are rather different from the STM case. At higher ${P_r}$ value, the DTM growth becomes slower, and the scales of the magnetic island, the plasmoid, and the flow vortices tend to be larger and more global. In future we plan on identifying the critical value of the ${P_r}$ quantitatively at which the generation of monster plasmoid becomes halted. Also we intend to explore the two-fluid and the 3D effects beyond this work next. 

\section*{Acknowledgement}

This research was supported by the National Magnetic Confinement Fusion Science Program of China (No.2019YFE03050004), National Natural Science Foundation of China (Grants Nos. 42130202, 41874190 and 51821005), U.S.DOE (Nos.DE-FG02-86ER53218 and DESC0018001), and the Fundamental Research Funds for the Central Universities at Huazhong University of Science and Technology (No.2019kfyXJJS193). We are grateful for the support from NIMROD team. This research used the computing resources from the Supercomputing Center of University of Science and Technology of China. The author Nisar Ahmad acknowledges the assistance from Dr. Haolong Li.
\section*{DATA AVAILABILITY}

The supporting data regarding the findings of this study is available
from the corresponding author upon reasonable request.

\newpage

%\section*{References}
\providecommand{\newblock}{}

%%%%%% end biblography =============

\begin{figure}[htbp]
\centering
\begin{minipage}[c]{0.5\linewidth}
\includegraphics[width=\textwidth]{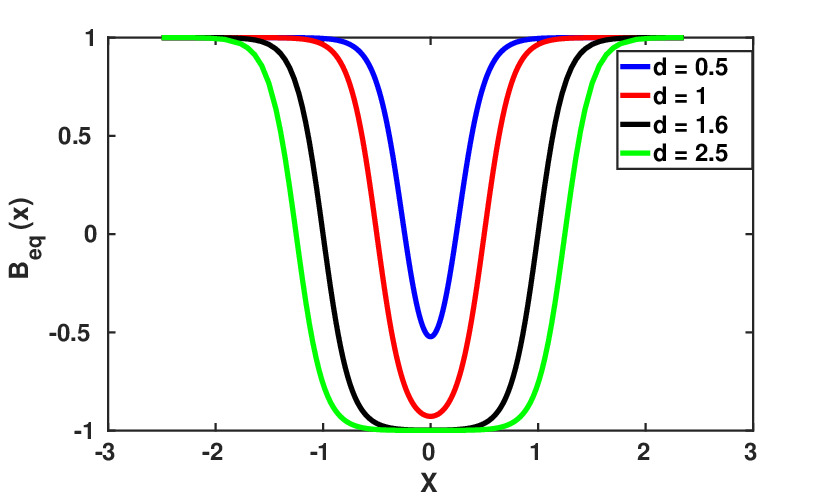}
\put(-250,165){\textbf{(a)}}
\end{minipage}
\begin{minipage}[c]{0.5\linewidth}
\includegraphics[width=\textwidth]{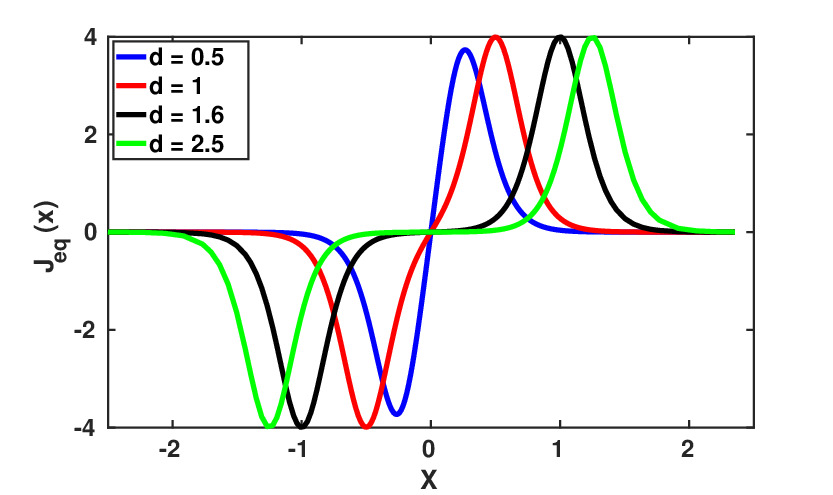}
\put(-250,165){\textbf{(b)}}
\end{minipage}
\begin{minipage}[c]{0.5\linewidth}
\includegraphics[width=\textwidth]{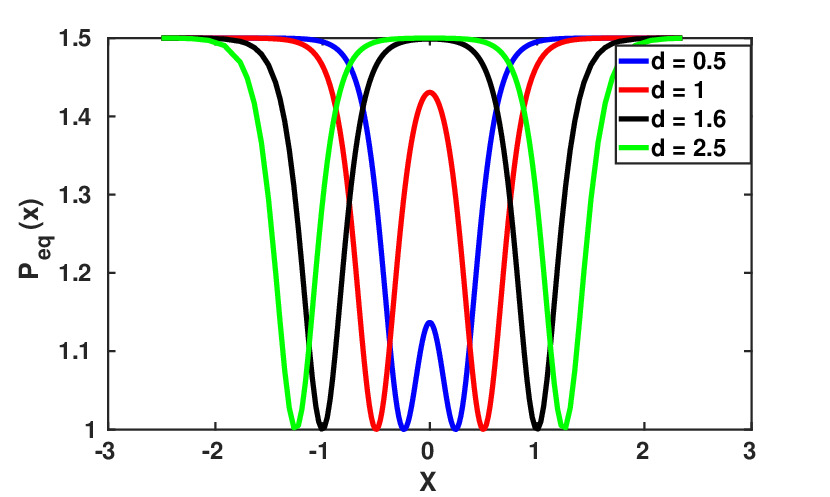}
\put(-250,165){\textbf{(c)}}
\end{minipage}
\caption{(a) Double Harris sheet equilibrium magnetic field, (b) current density and (c) pressure profiles at different distances between the rational surfaces.}
\end{figure}

\begin{figure}[htbp]
\centering
\begin{minipage}{0.43\textwidth}
\includegraphics[width=1\textwidth]{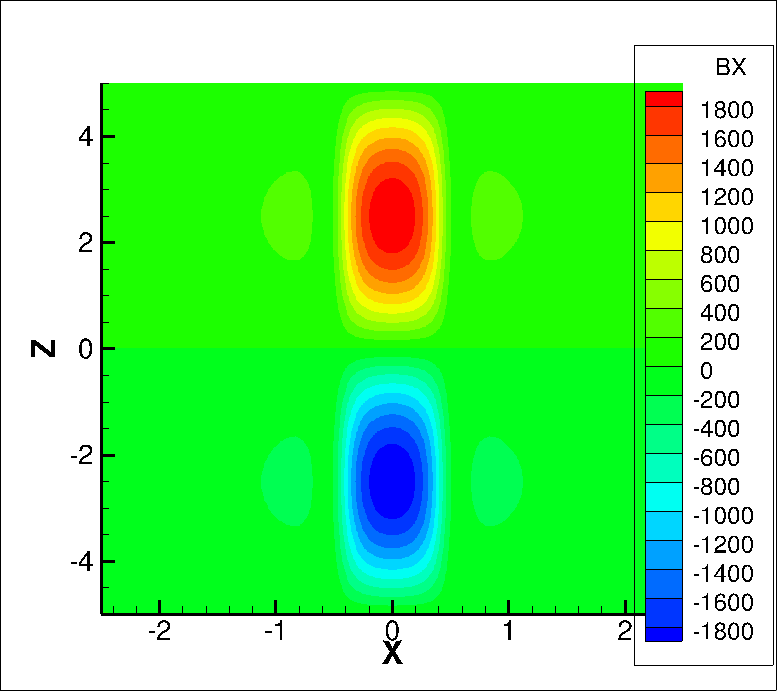}
\put(-170,160){\textbf{(a)}}
\end{minipage}
\begin{minipage}{0.43\textwidth}
\includegraphics[width=1\textwidth]{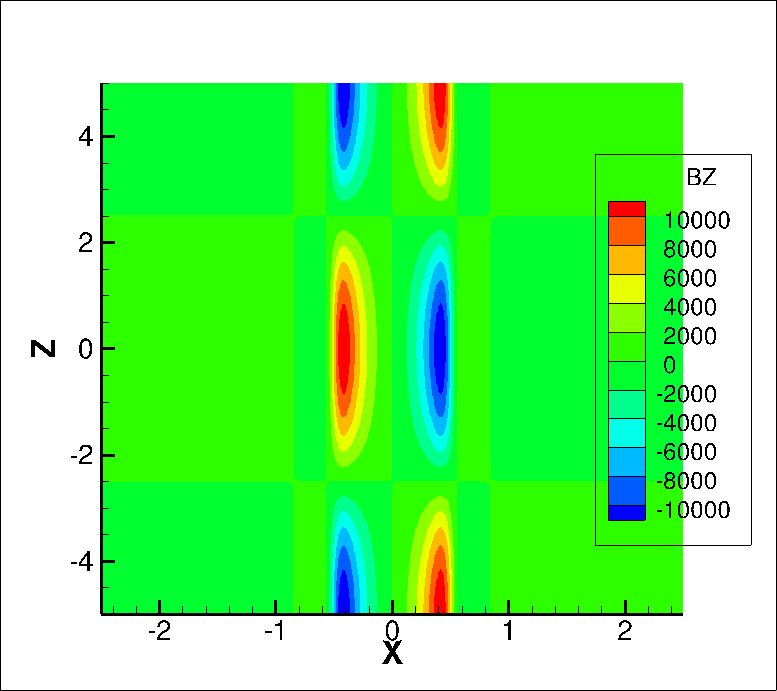}
\put(-170,160){\textbf{(b)}}
\end{minipage}
\begin{minipage}{0.43\textwidth}
\includegraphics[width=1\textwidth]{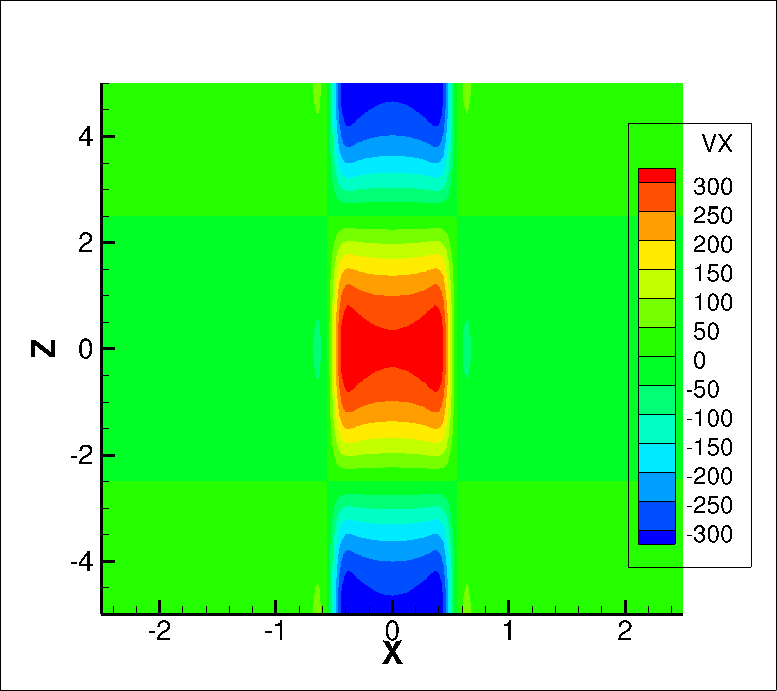}
\put(-170,160){\textbf{(c)}}
\end{minipage}
\begin{minipage}{0.43\textwidth}
\includegraphics[width=1\textwidth]{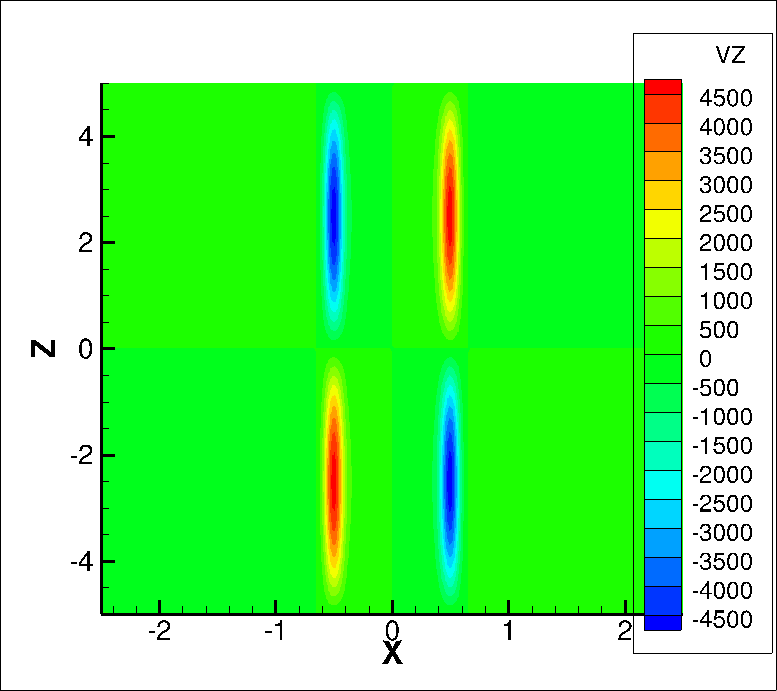}
\put(-170,160){\textbf{(d)}}
\end{minipage}
\begin{minipage}{0.43\textwidth}
\includegraphics[width=1\textwidth]{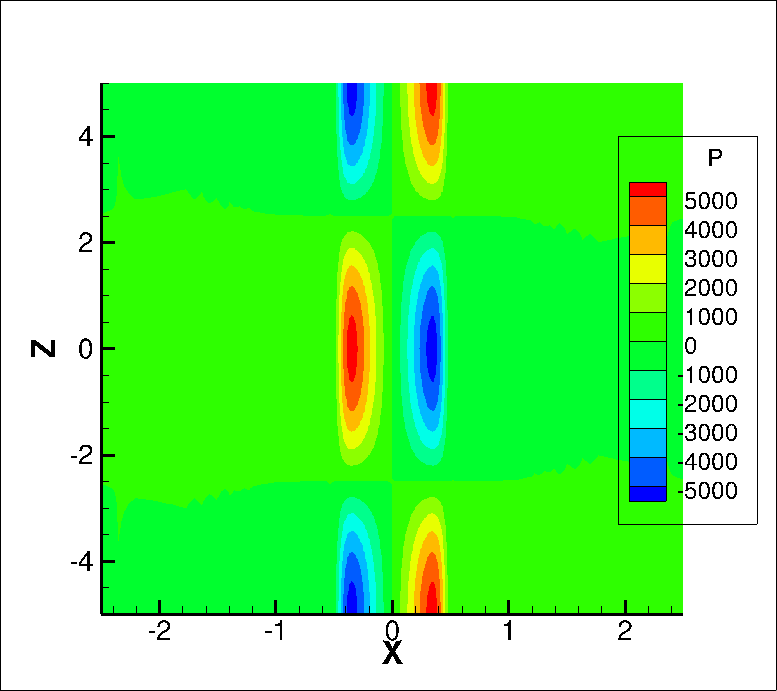}
\put(-170,160){\textbf{(e)}}
\end{minipage}
\caption{Linear mode structures of (a) magnetic field component ${B_x}$, (b) magnetic field component ${B_z}$, (c) velocity component ${v_x}$, (d) velocity component ${v_z}$ and (e) pressure ${P}$ at ${ d = 1}$.}
\end{figure}

\begin{figure}[htbp]
\centering
\begin{minipage}[c]{\textwidth}
\includegraphics[width=1.0\textwidth]{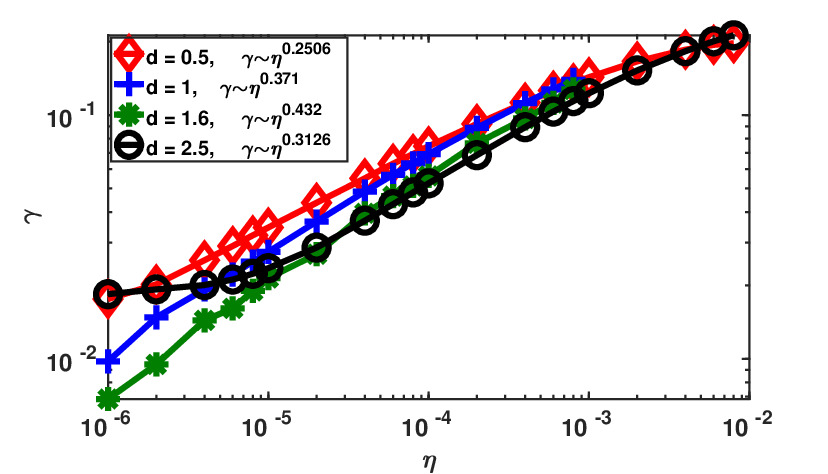}
\put(-400,255){\textbf{(a)}}
\end{minipage}
\begin{minipage}[c]{\textwidth}
\includegraphics[width=1.0\textwidth]{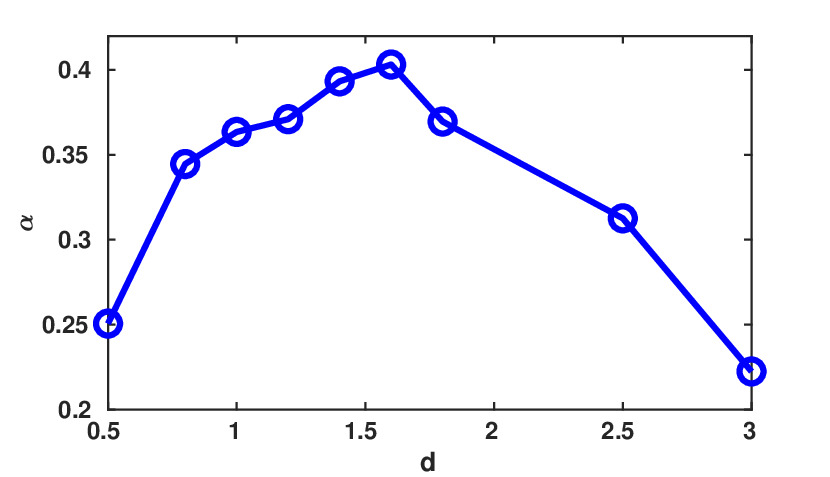}
\put(-400,255){\textbf{(b)}}
\end{minipage}
\caption{ (a)Linear growth rate as functions of the resistivity for different distances between the rational surfaces (b) ${\alpha}$ (from $\gamma_{lin}$ = ${\eta^{\alpha}}$ ) is function of the distances between the rational surfaces.}
\end{figure}

\begin{figure}[htbp]
\centering
\begin{minipage}[c]{\textwidth}
\includegraphics[width=1\textwidth]{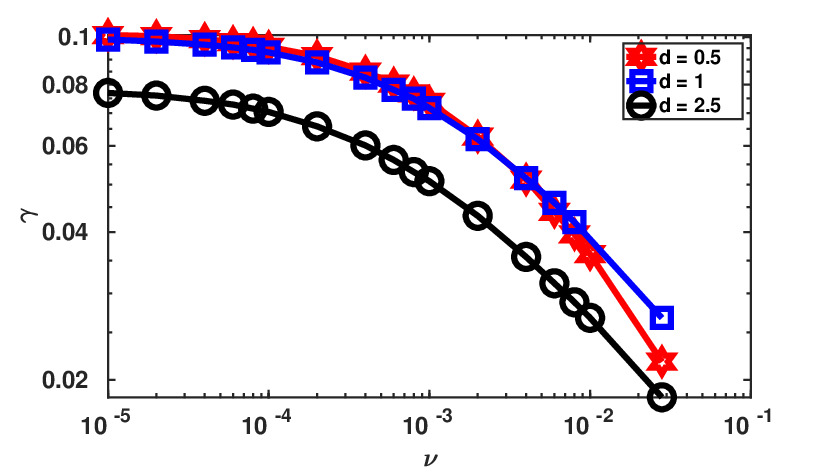}
\put(-400,255){\textbf{(a)}}
\end{minipage}
\begin{minipage}[c]{\textwidth}
\includegraphics[width=1\textwidth]{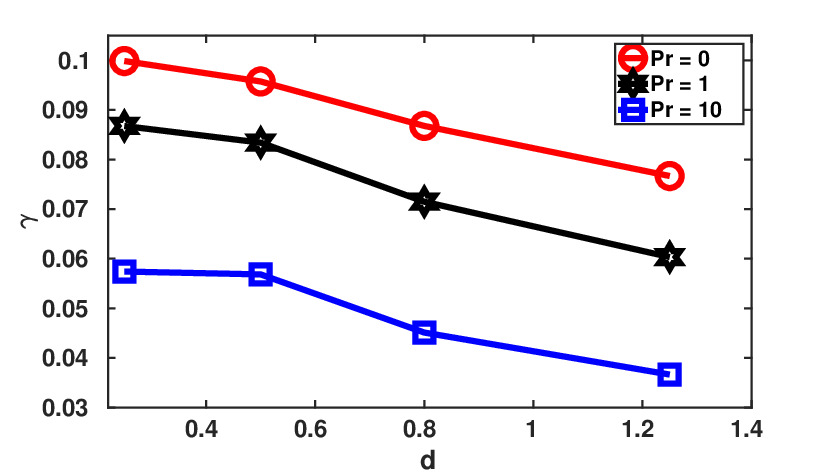}
\put(-400,255){\textbf{(b)}}
\end{minipage}
\caption{ (a) Linear growth rate as functions of the viscosity at different distances between the rational surfaces (b) Linear growth rate as functions of the distances between the rational surfaces for different Prandtl numbers.}
\end{figure}

\begin{figure}
 \includegraphics[width=1.1\linewidth]{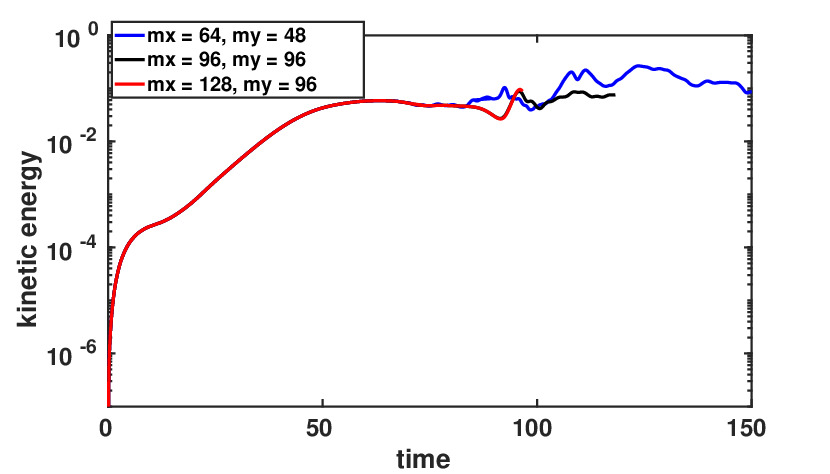}\\
 \caption{Kinetic energy evolution for different numerical resolutions at ${d = 2(x_0) = 1.1}$.}
\end{figure} 

\begin{figure}[htbp]
\centering
\begin{minipage}{0.425\textwidth}
\includegraphics[width=1.0\textwidth]{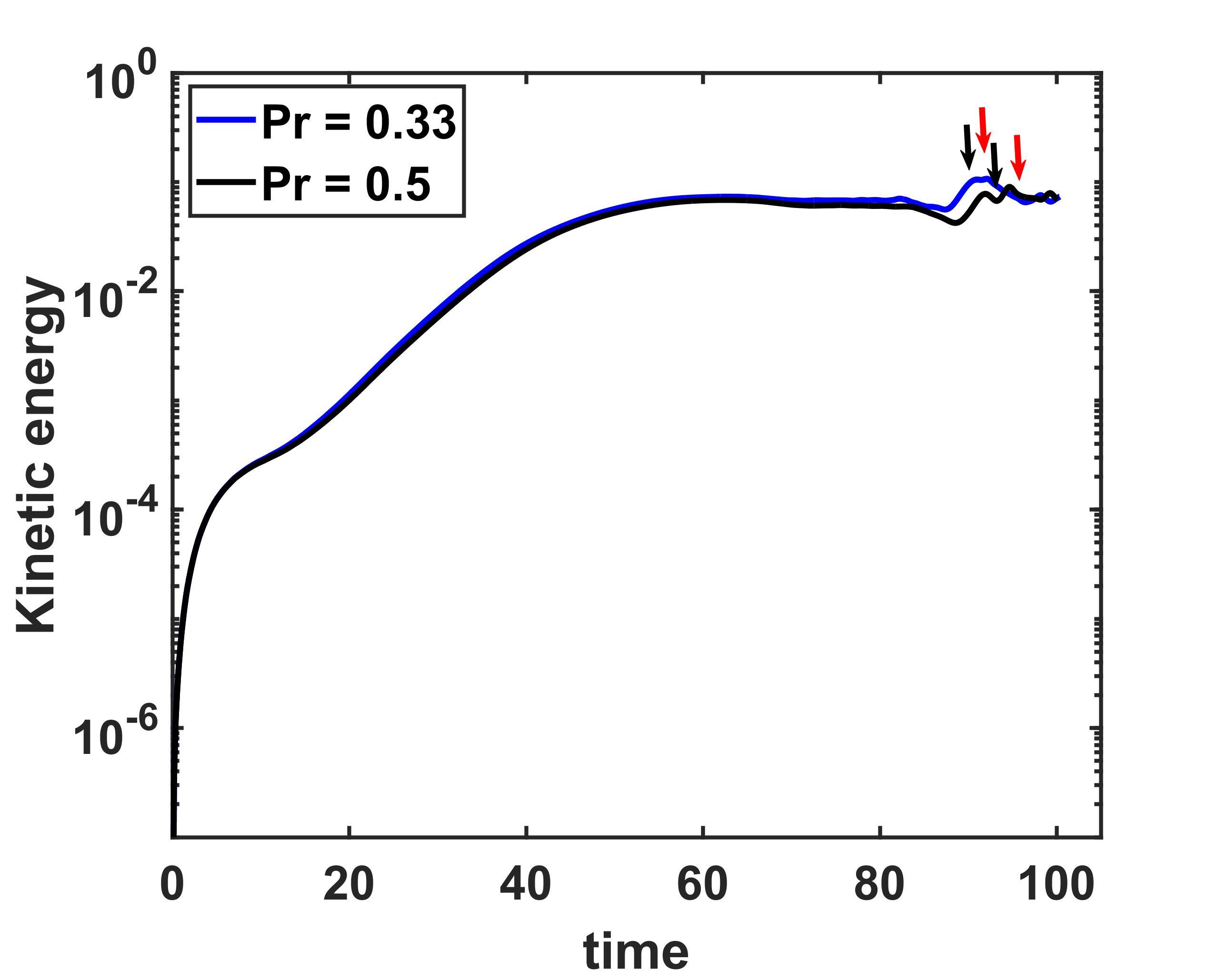}
\put(-175,155){\textbf{(a)}}
\end{minipage}
\begin{minipage}{0.45\textwidth}
\includegraphics[width=1.0\textwidth]{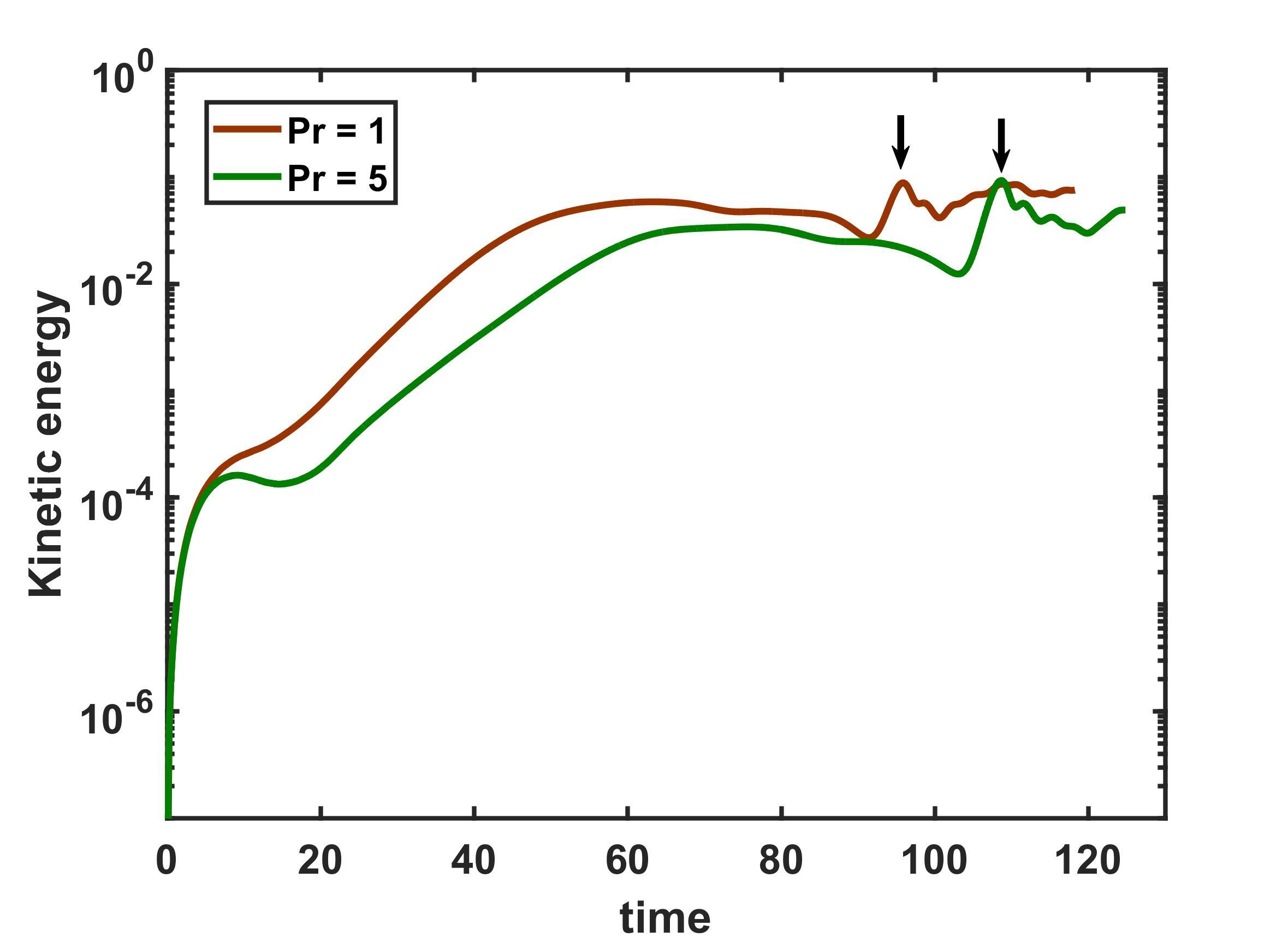}
\put(-180,150){\textbf{(b)}}
\end{minipage}
\begin{minipage}{0.45\textwidth}
\includegraphics[width=1.0\textwidth]{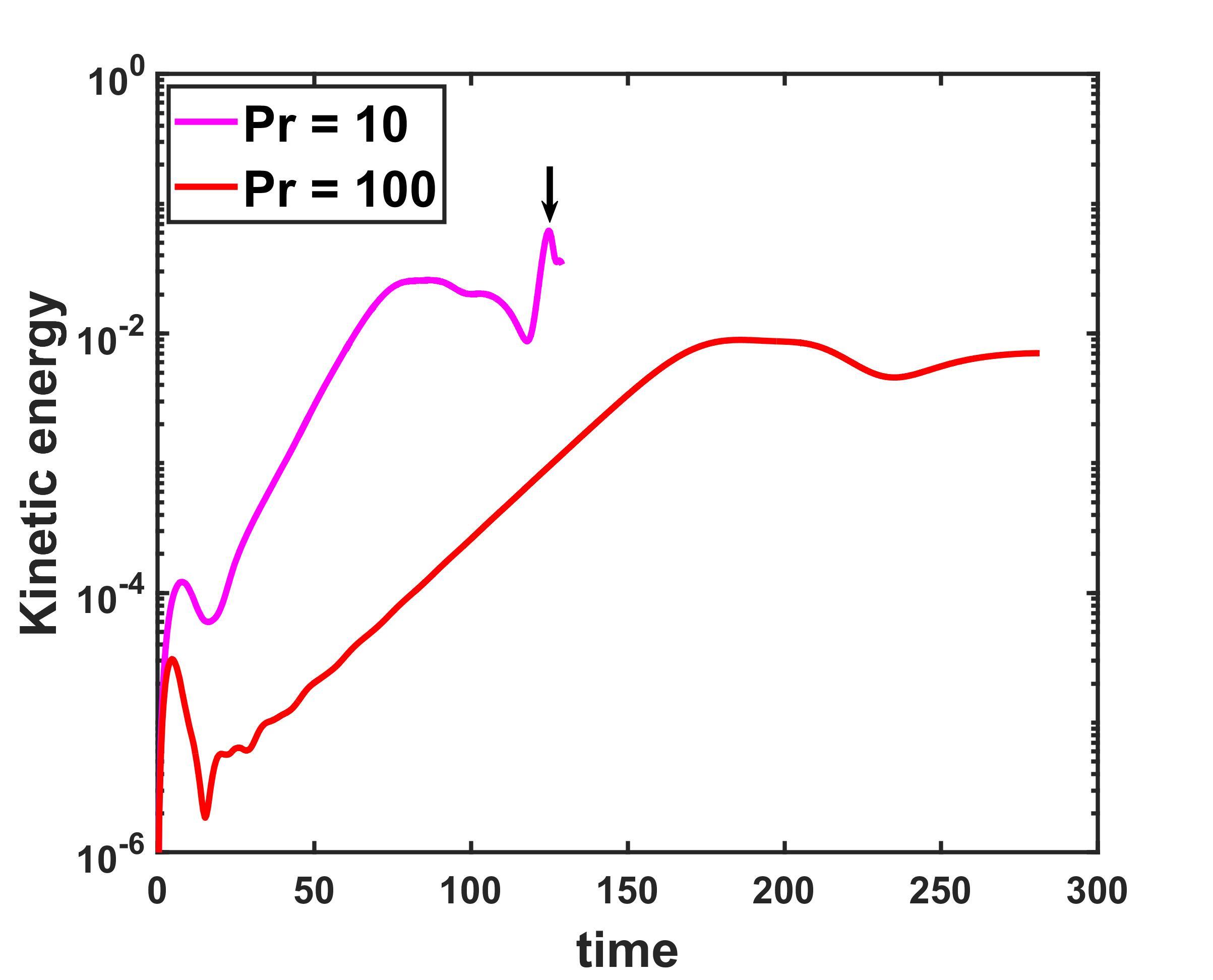}
\put(-180,160){\textbf{(c)}}
\end{minipage}
 \caption{Kinetic energy evolution for (a) ${P_r = 0.33, 0.5}$, (b) ${P_r = 1,5}$ and (c) ${P_r = 10, 100}$ at ${d = 1.1}$.}
\end{figure} 

\begin{figure}[htbp]
\centering
\begin{minipage}{0.38\textwidth}
\includegraphics[width=1.0\textwidth]{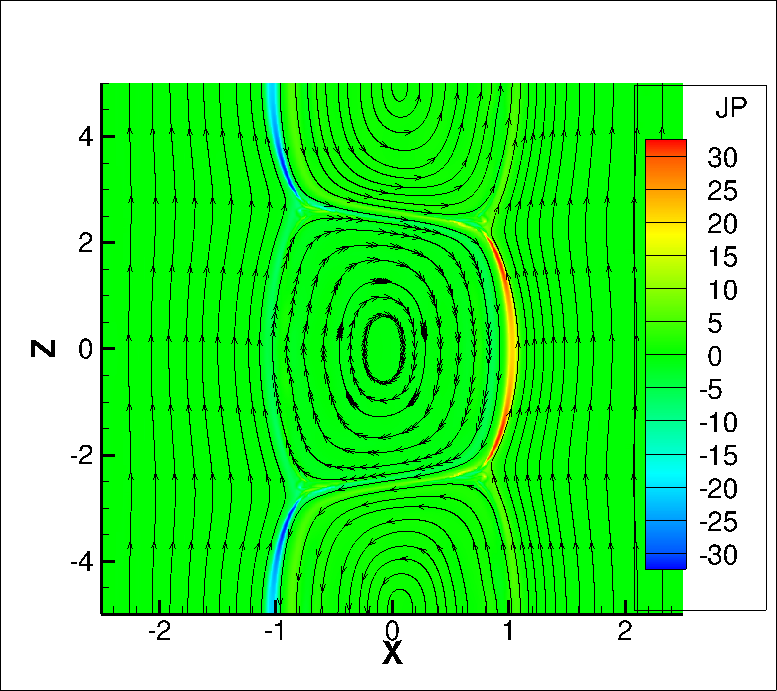}
\put(-185,160){\textbf{(a)}}
\end{minipage}
\begin{minipage}{0.38\textwidth}
\includegraphics[width=1.0\textwidth]{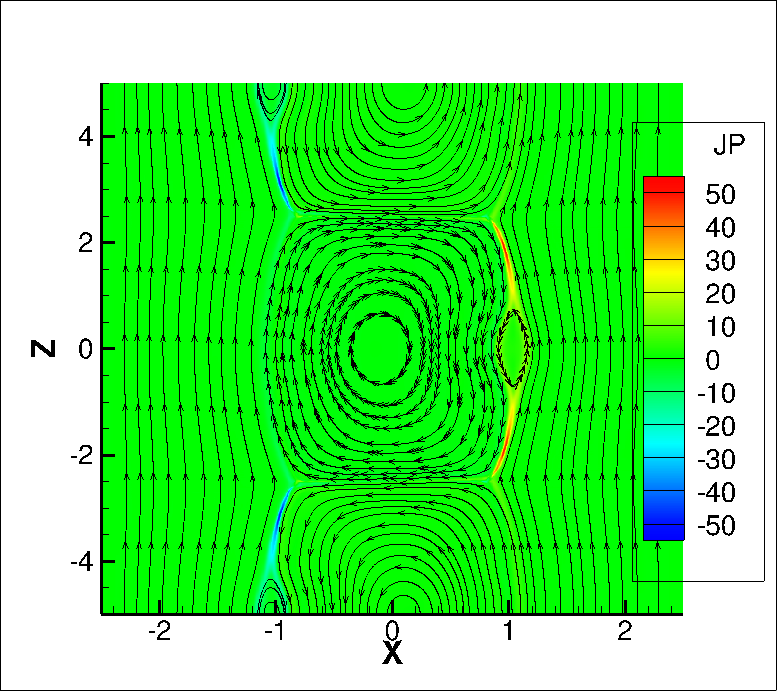}
\put(-185,160){\textbf{(b)}}
\end{minipage}
\begin{minipage}{0.38\textwidth}
\includegraphics[width=1.0\textwidth]{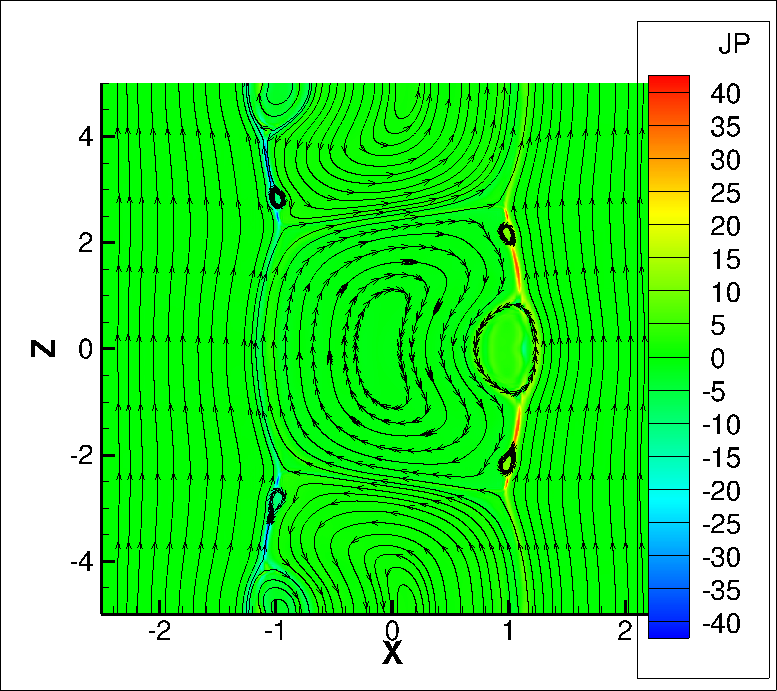}
\put(-185,160){\textbf{(c)}}
\end{minipage}
\begin{minipage}{0.38\textwidth}
\includegraphics[width=1.0\textwidth]{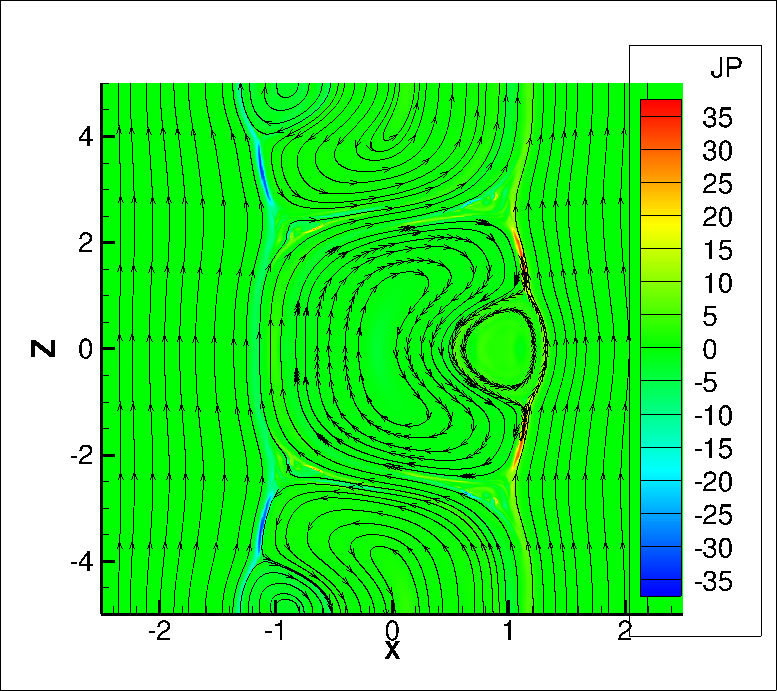}
\put(-185,160){\textbf{(d)}}
\end{minipage}
\caption{2D contours of the current density in z-direction and the 2D magnetic field lines at time (a) t= 85, (b) t = 90, (c) t = 95 and (d) t = 100 with Pr =0.33 and ${d = 1.1}$.}
\end{figure}

\begin{figure}[htbp]
\centering
\begin{minipage}{0.38\textwidth}
\includegraphics[width=1.0\textwidth]{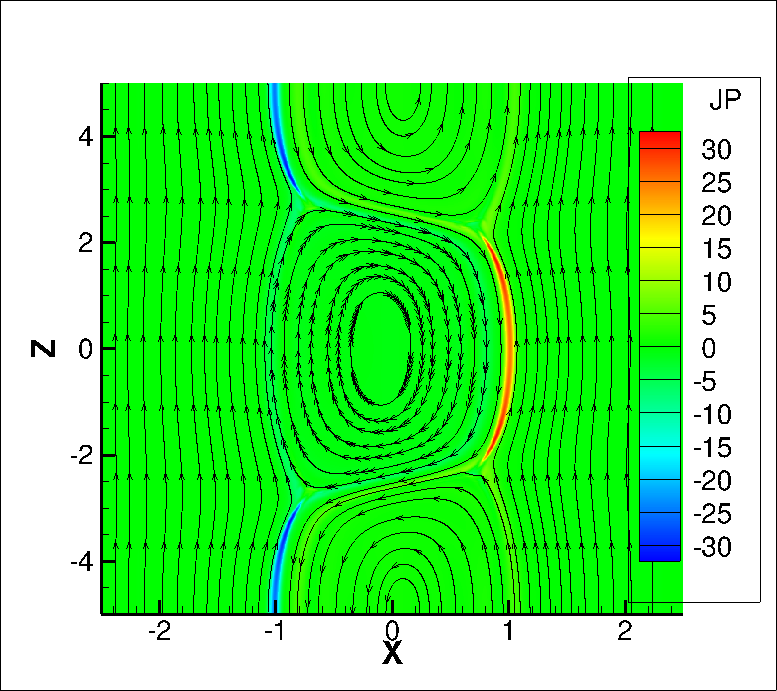}
\put(-185,160){\textbf{(a)}}
\end{minipage}
\begin{minipage}{0.38\textwidth}
\includegraphics[width=1.0\textwidth]{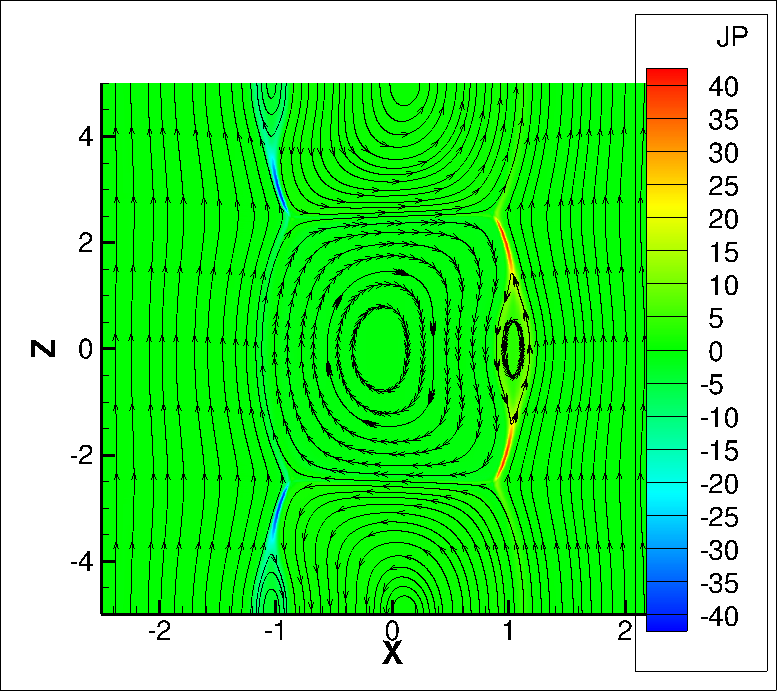}
\put(-185,160){\textbf{(b)}}
\end{minipage}
\begin{minipage}{0.38\textwidth}
\includegraphics[width=1.0\textwidth]{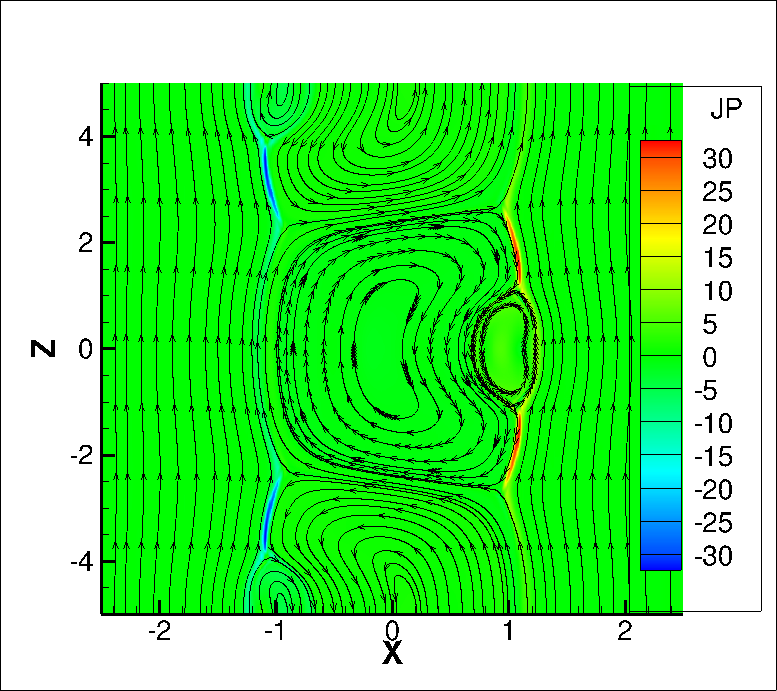}
\put(-185,160){\textbf{(c)}}
\end{minipage}
\begin{minipage}{0.38\textwidth}
\includegraphics[width=1.0\textwidth]{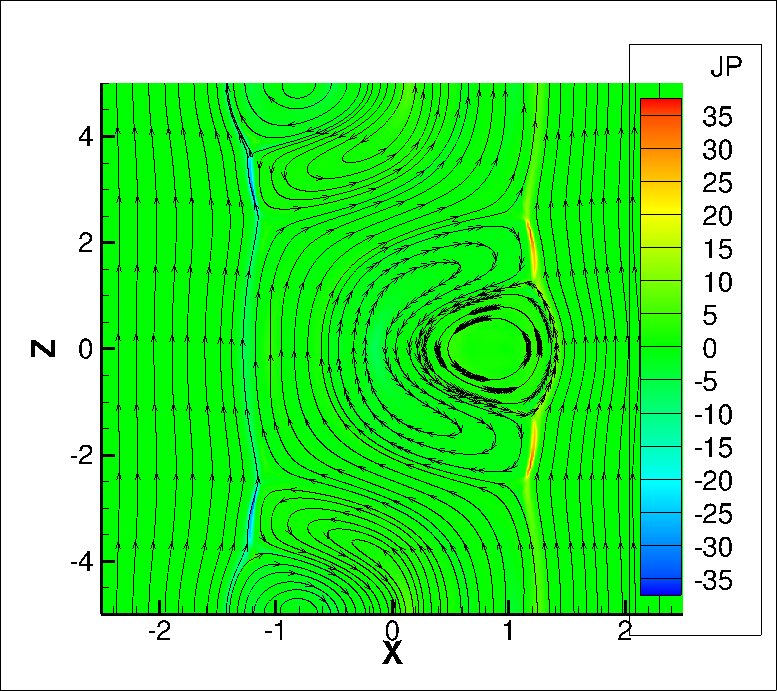}
\put(-185,160){\textbf{(d)}}
\end{minipage}
\caption{2D contours of the current density in z-direction and the 2D magnetic field lines at  time (a) t = 85, (b) t = 95, (c) t = 100 and (d) t = 115 with Pr = 1 and ${d = 1.1}$.}
\end{figure}

\begin{figure}[htbp]
\centering
\begin{minipage}{0.38\textwidth}
\includegraphics[width=1.0\textwidth]{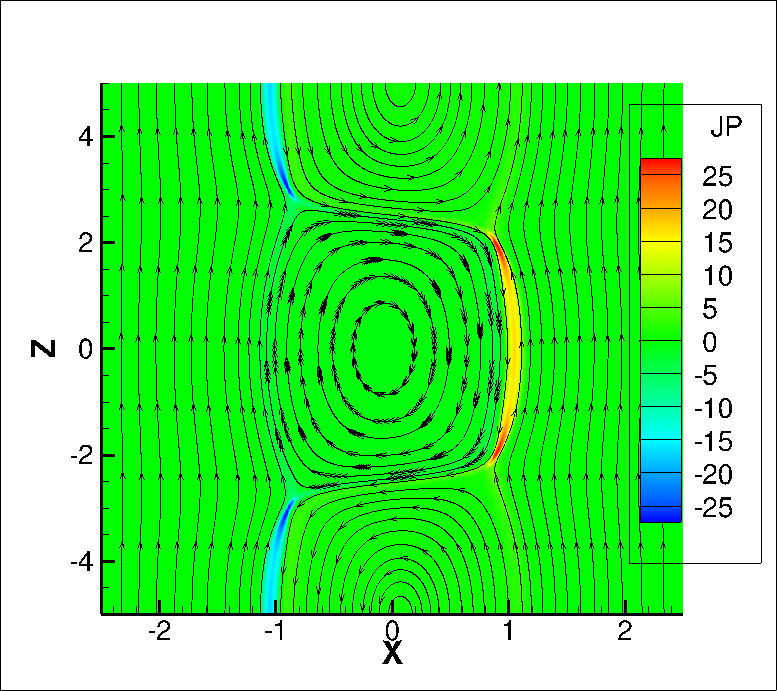}
\put(-185,160){\textbf{(a)}}
\end{minipage}
\begin{minipage}{0.38\textwidth}
\includegraphics[width=1.0\textwidth]{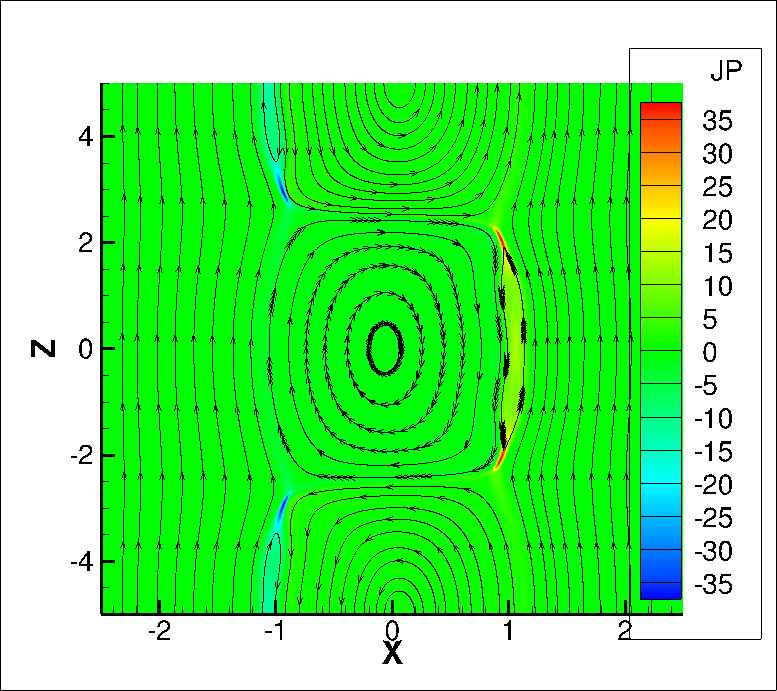}
\put(-185,160){\textbf{(b)}}
\end{minipage}
\begin{minipage}{0.38\textwidth}
\includegraphics[width=1.0\textwidth]{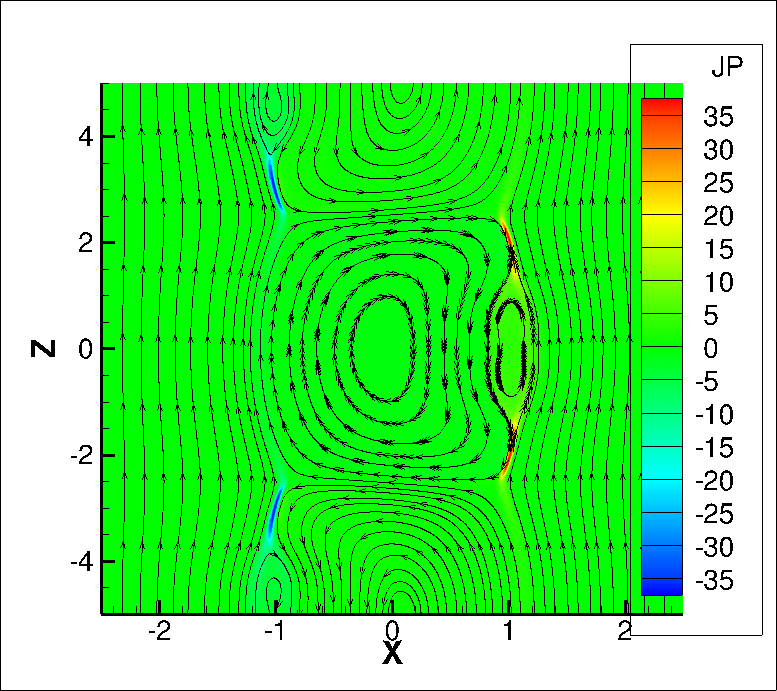}
\put(-185,160){\textbf{(c)}}
\end{minipage}
\begin{minipage}{0.38\textwidth}
\includegraphics[width=1.0\textwidth]{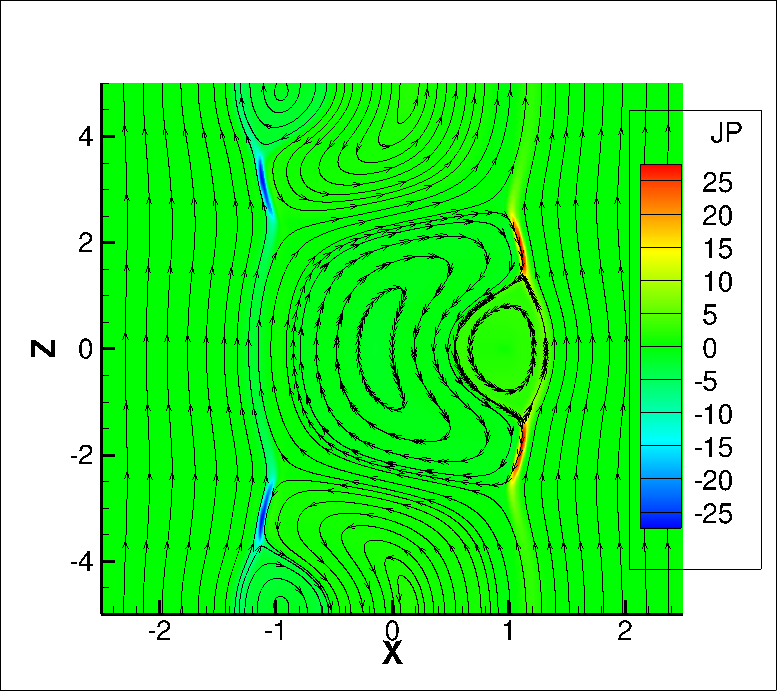}
\put(-185,160){\textbf{(d)}}
\end{minipage}
\caption{2D contours of the current density in z-direction and the 2D magnetic field lines at time (a) t = 116, (b) t = 120, (c) t = 124 and (d) t = 132 with Pr = 10 and ${d = 1.1}$.}
\end{figure}

\begin{figure}[htbp]
\centering
\begin{minipage}{0.38\textwidth}
\includegraphics[width=1.0\textwidth]{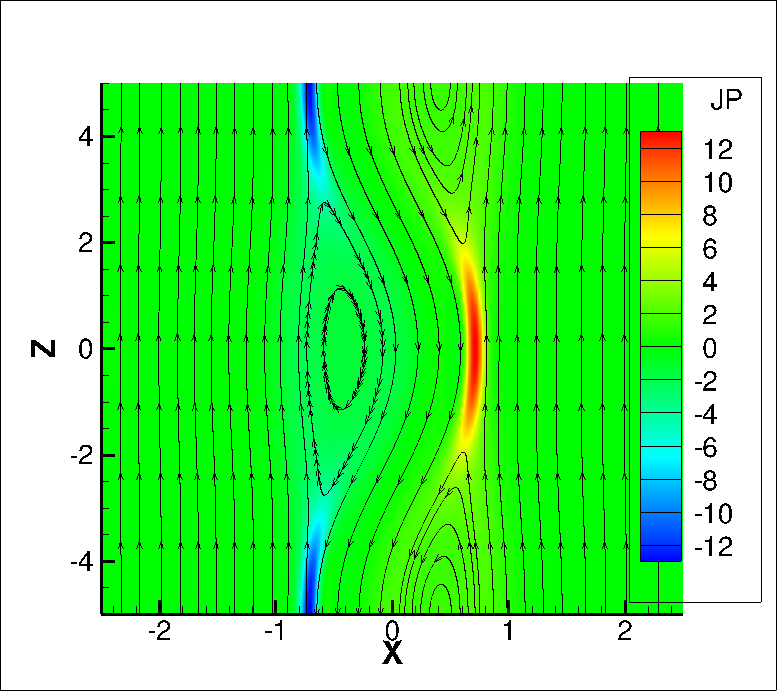}
\put(-185,160){\textbf{(a)}}
\end{minipage}
\begin{minipage}{0.38\textwidth}
\includegraphics[width=1.0\textwidth]{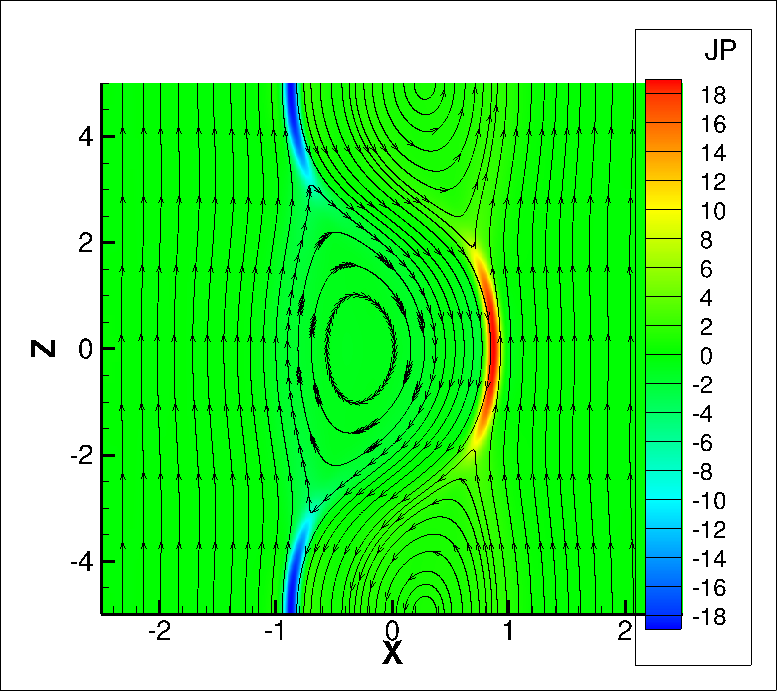}
\put(-185,160){\textbf{(b)}}
\end{minipage}
\begin{minipage}{0.38\textwidth}
\includegraphics[width=1.0\textwidth]{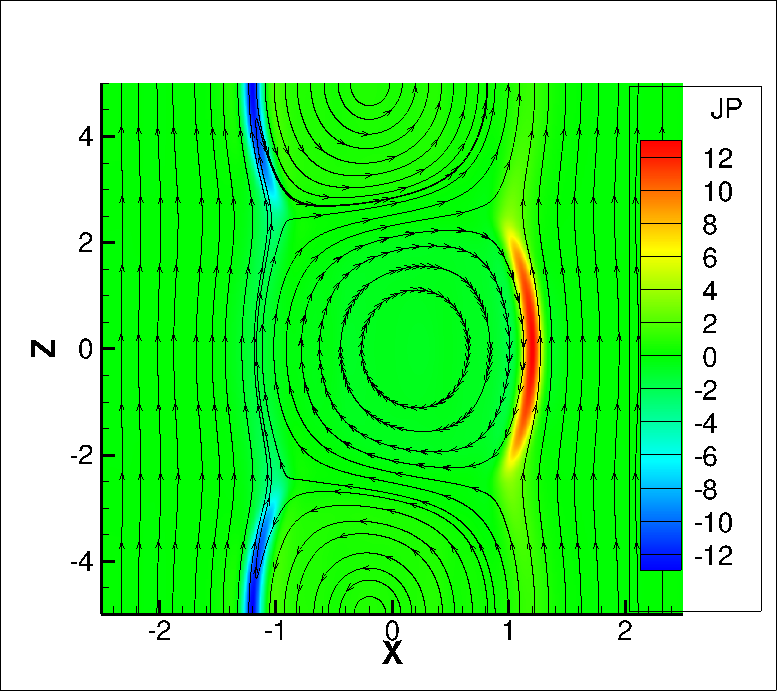}
\put(-185,160){\textbf{(c)}}
\end{minipage}
\begin{minipage}{0.38\textwidth}
\includegraphics[width=1.0\textwidth]{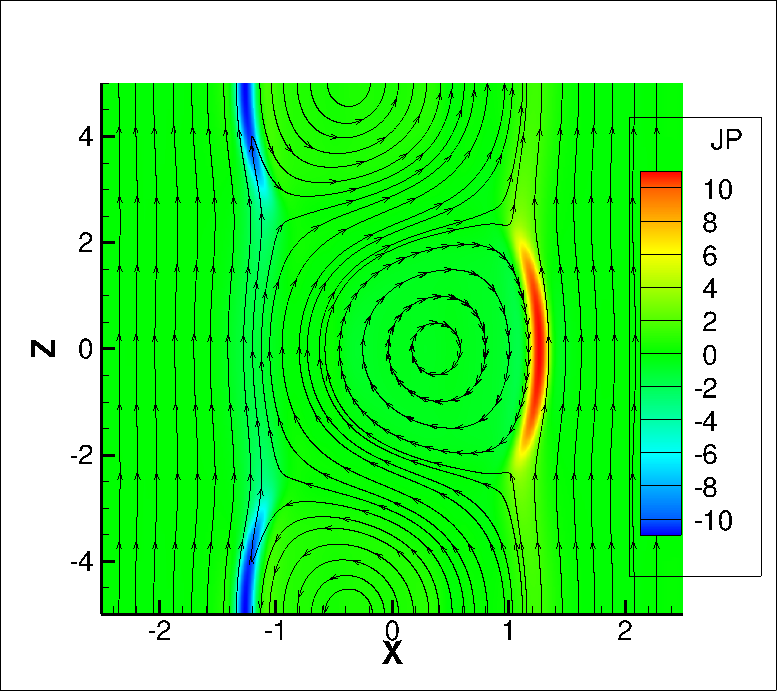}
\put(-185,160){\textbf{(d)}}
\end{minipage}
\caption{2D contours of the current density in z-direction and the 2D magnetic field lines at time (a) t = 160, (b) t = 185, (c) t = 250 and (d) t = 270 with Pr = 100 and ${d = 1.1}$.}
\end{figure}

\begin{figure}[htbp]
\centering
\begin{minipage}{1.0\textwidth}
\includegraphics[width=1.0\textwidth]{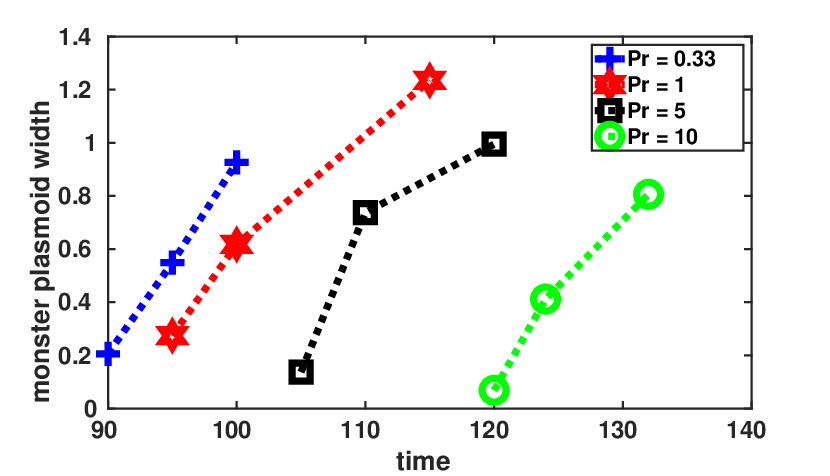}
\put(-400,260){\textbf{(a)}}
\end{minipage}
\begin{minipage}{1.0\textwidth}
\includegraphics[width=1.0\textwidth]{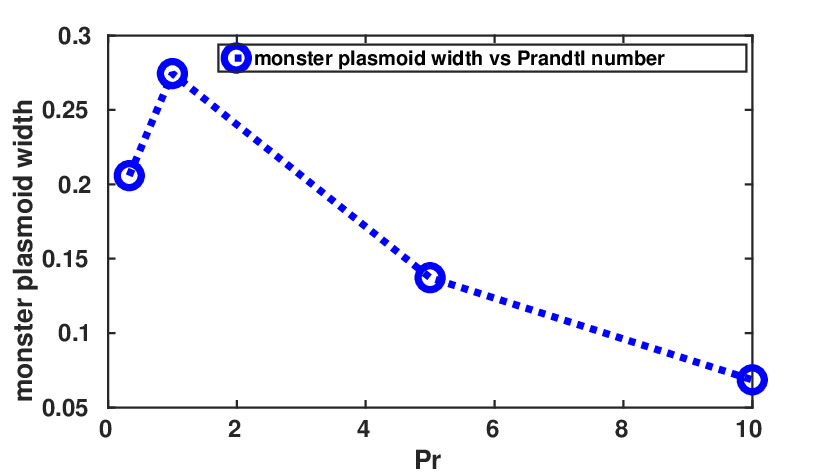}
\put(-400,260){\textbf{(b)}}
\end{minipage}
\caption{(a) Monster plasmoid evolution (b) Monster plasmoid width as a function of ${P_r}$ with ${P_r = 0.33, 1, 5}$ and ${10}$.}
\end{figure}

\begin{figure}[htbp]
\centering
\begin{minipage}{0.38\textwidth}
\includegraphics[width=1.0\textwidth]{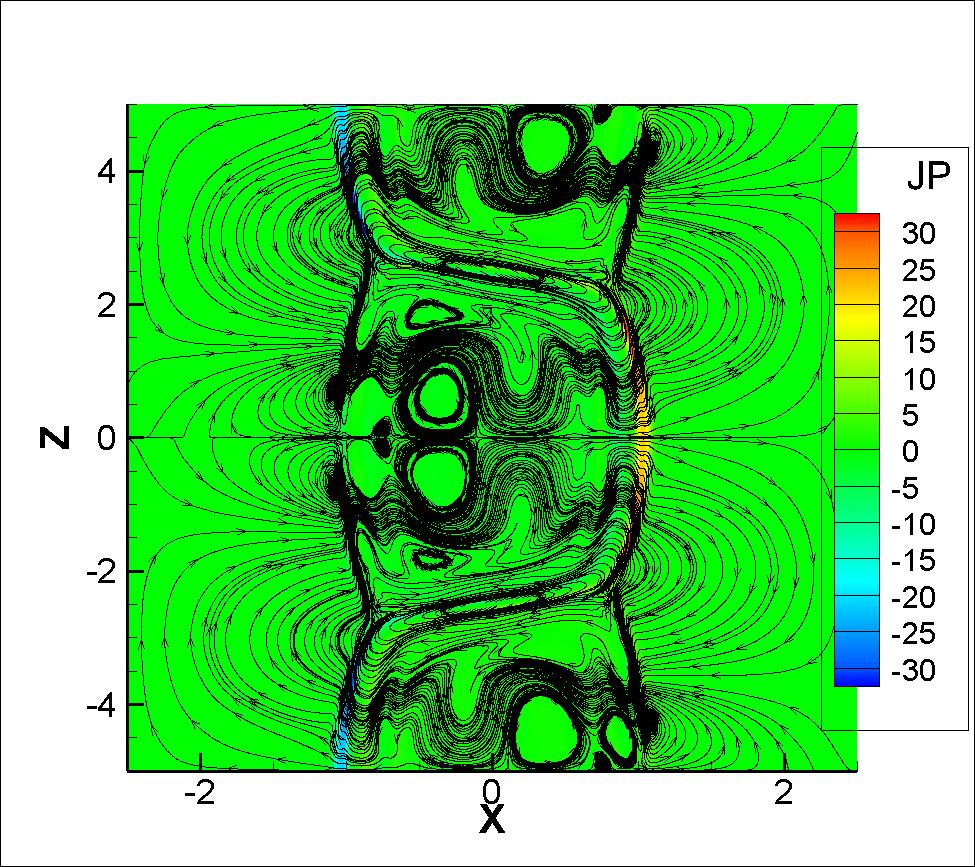}
\put(-175,140){\textbf{(a) t = 85}}
\end{minipage}
\begin{minipage}{0.38\textwidth}
\includegraphics[width=1.0\textwidth]{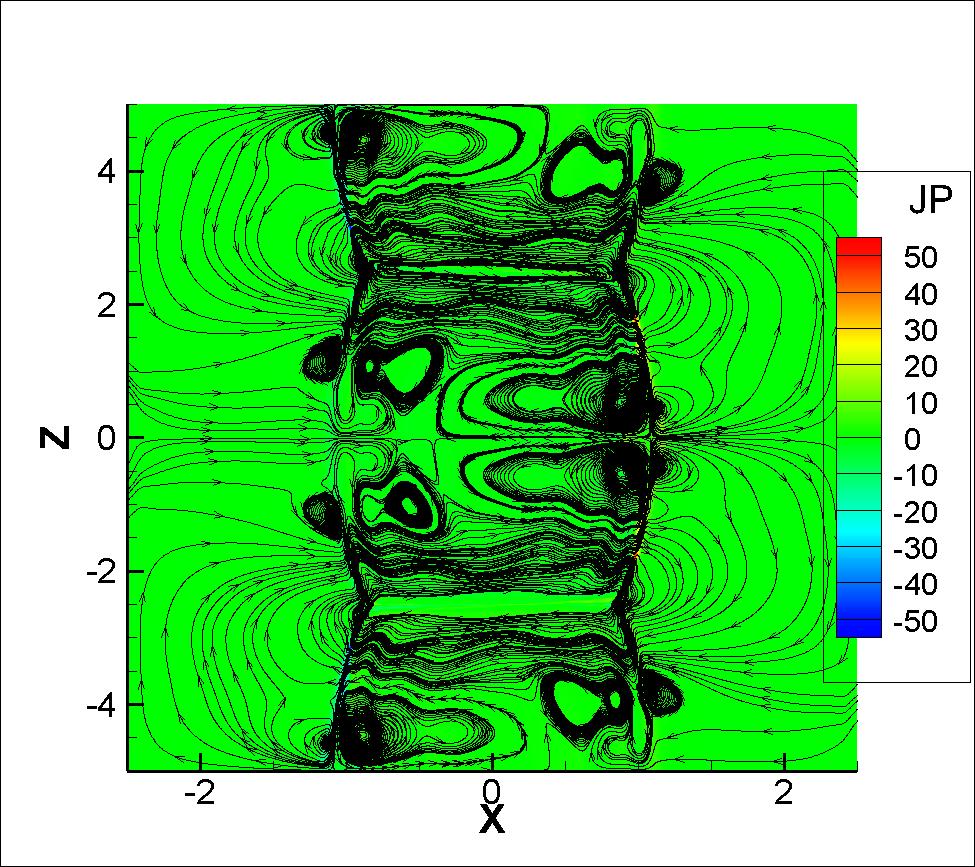}
\put(-175,140){\textbf{(b) t = 90}}
\end{minipage}
\begin{minipage}{0.38\textwidth}
\includegraphics[width=1.0\textwidth]{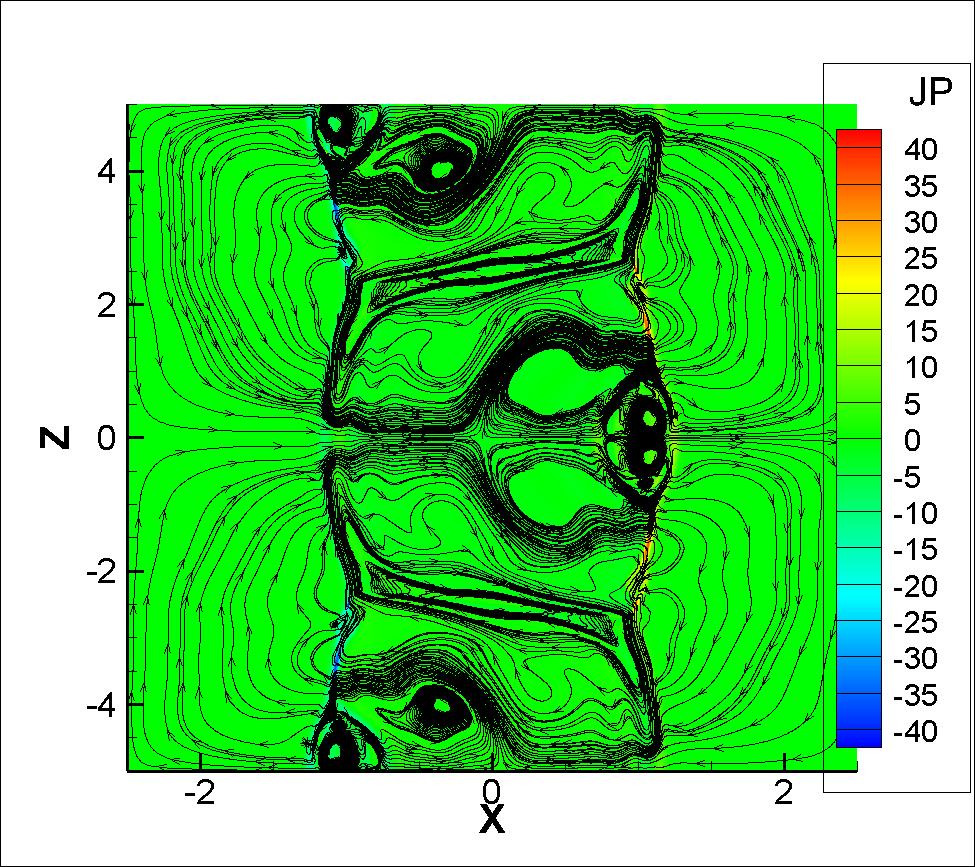}
\put(-175,140){\textbf{(c) t = 95}}
\end{minipage}
\begin{minipage}{0.38\textwidth}
\includegraphics[width=1.0\textwidth]{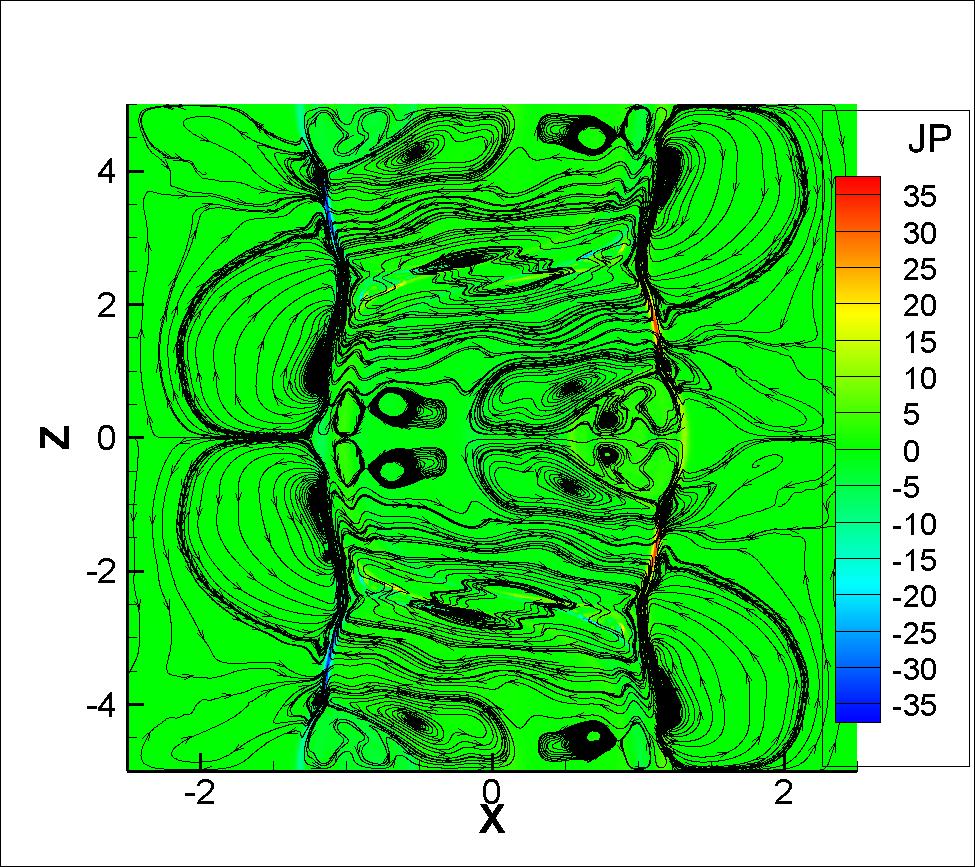}
\put(-175,140){\textbf{(d) t = 100}}
\end{minipage}
\begin{minipage}{0.38\textwidth}
\includegraphics[width=1.0\textwidth]{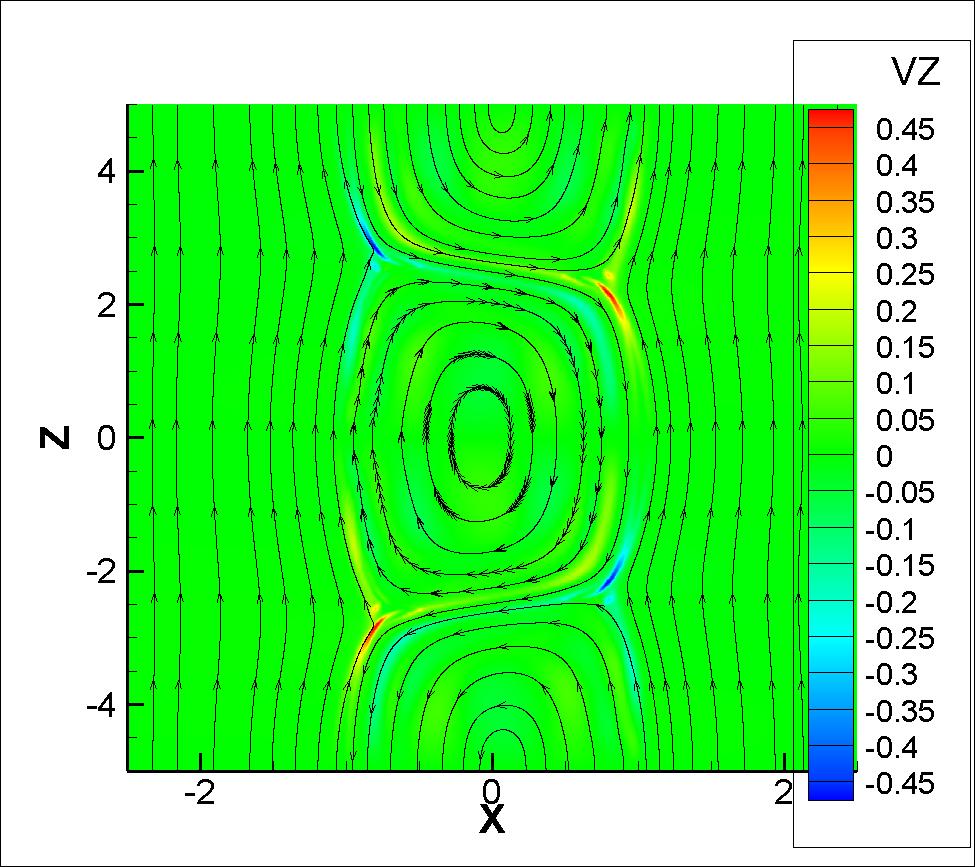}
\put(-175,140){\textbf{(e) t = 85}}
\end{minipage}
\begin{minipage}{0.38\textwidth}
\includegraphics[width=1.0\textwidth]{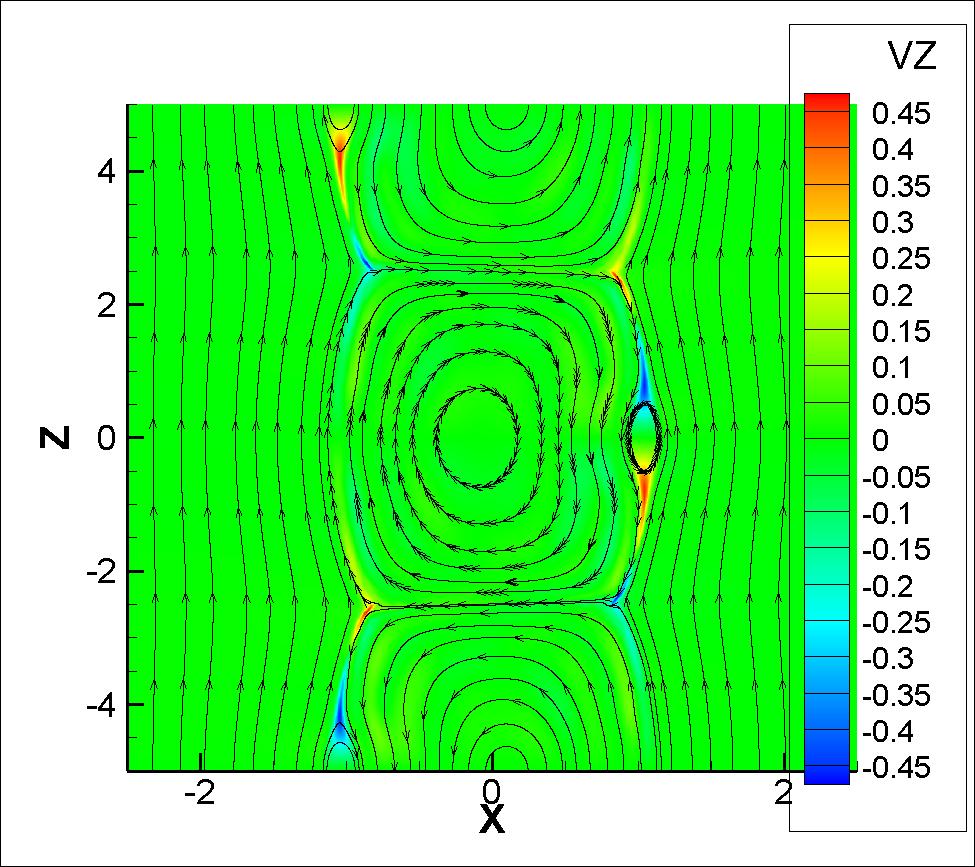}
\put(-175,140){\textbf{(f) t = 90}}
\end{minipage}
\begin{minipage}{0.38\textwidth}
\includegraphics[width=1.0\textwidth]{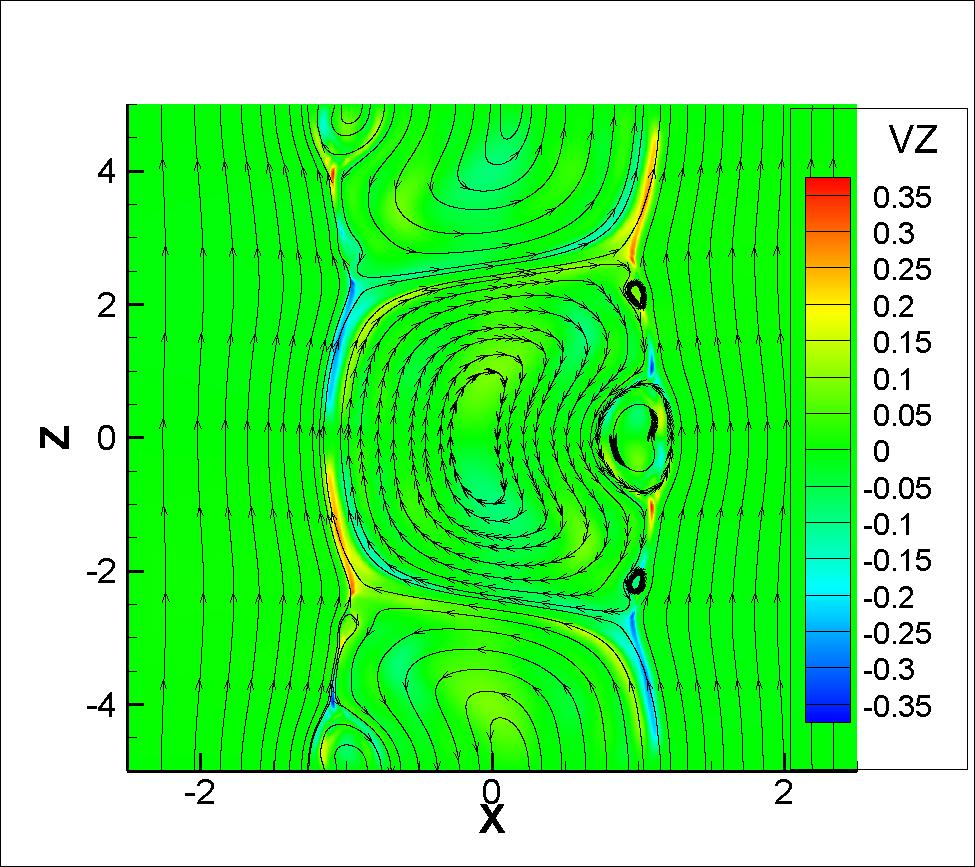}
\put(-175,140){\textbf{(g) t = 95}}
\end{minipage}
\begin{minipage}{0.38\textwidth}
\includegraphics[width=1.0\textwidth]{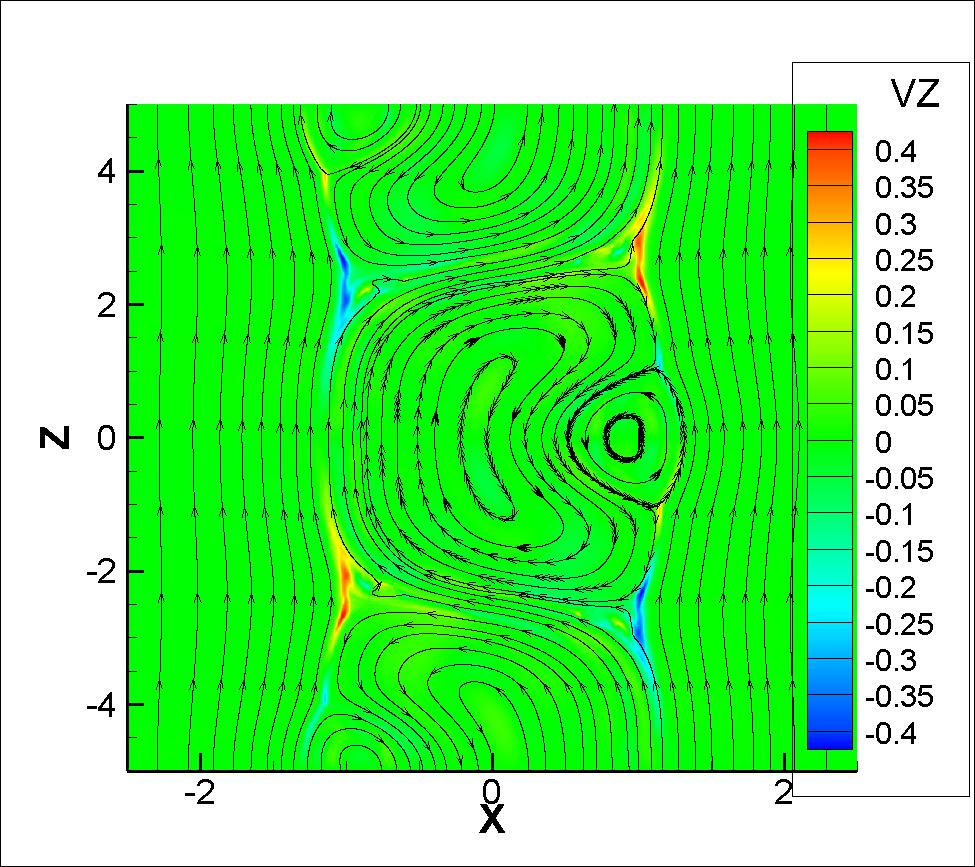}
\put(-175,140){\textbf{(h) t = 100}}
\end{minipage}
\caption{2D contours of the current density in z-direction and flow field stream lines (a-d), 2D contours of the flow field in z-direction and 2D magnetic field lines (e-h), Pr = 0.33 with ${d = 1.1}$.}
\end{figure}

\begin{figure}[htbp]
\centering
\begin{minipage}{0.38\textwidth}
\includegraphics[width=1.0\textwidth]{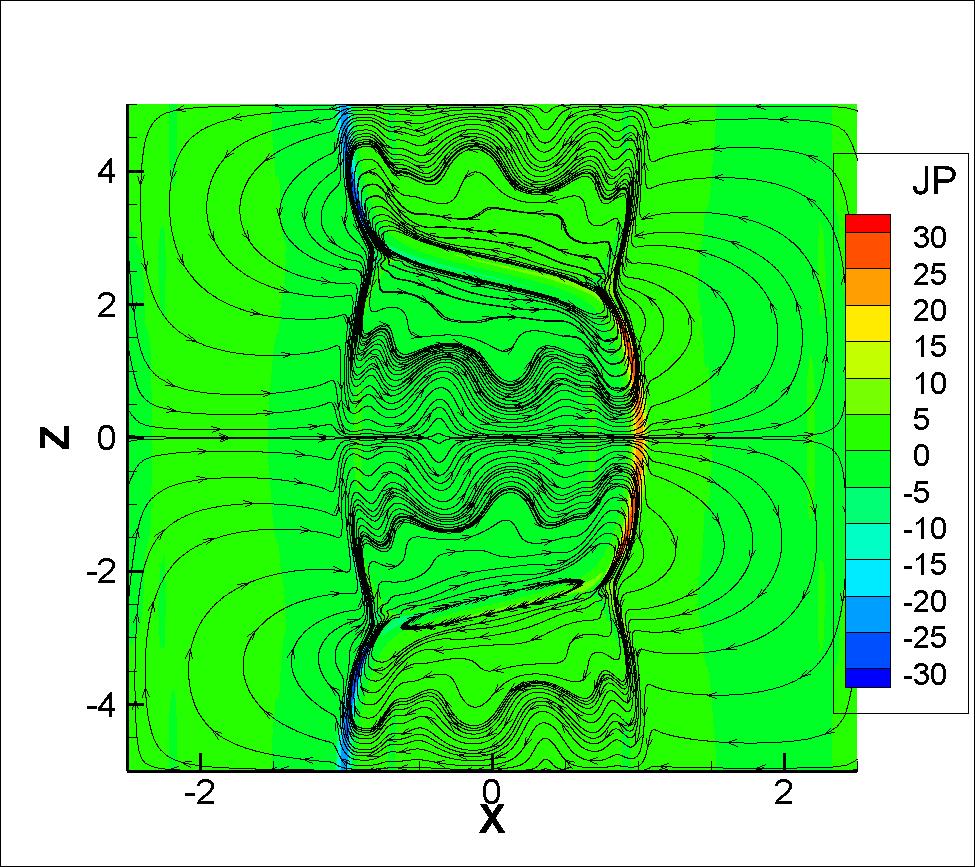}
\put(-175,140){\textbf{(a) t = 85}}
\end{minipage}
\begin{minipage}{0.38\textwidth}
\includegraphics[width=1.0\textwidth]{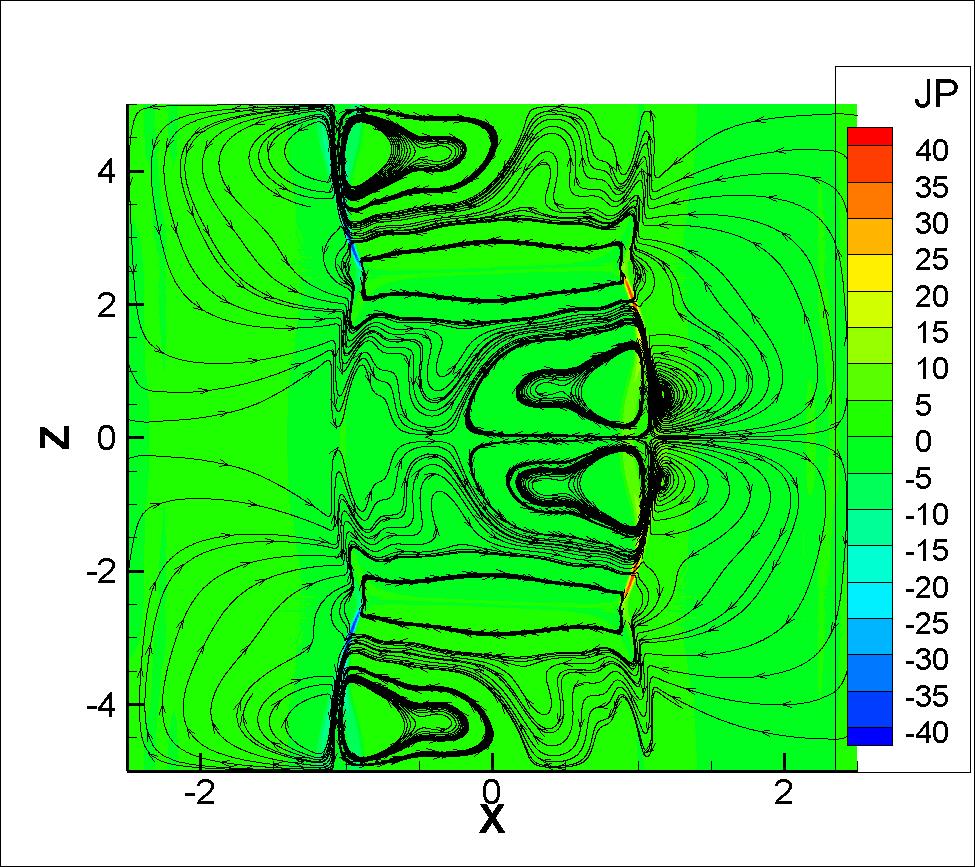}
\put(-175,140){\textbf{(b) t = 95}}
\end{minipage}
\begin{minipage}{0.38\textwidth}
\includegraphics[width=1.0\textwidth]{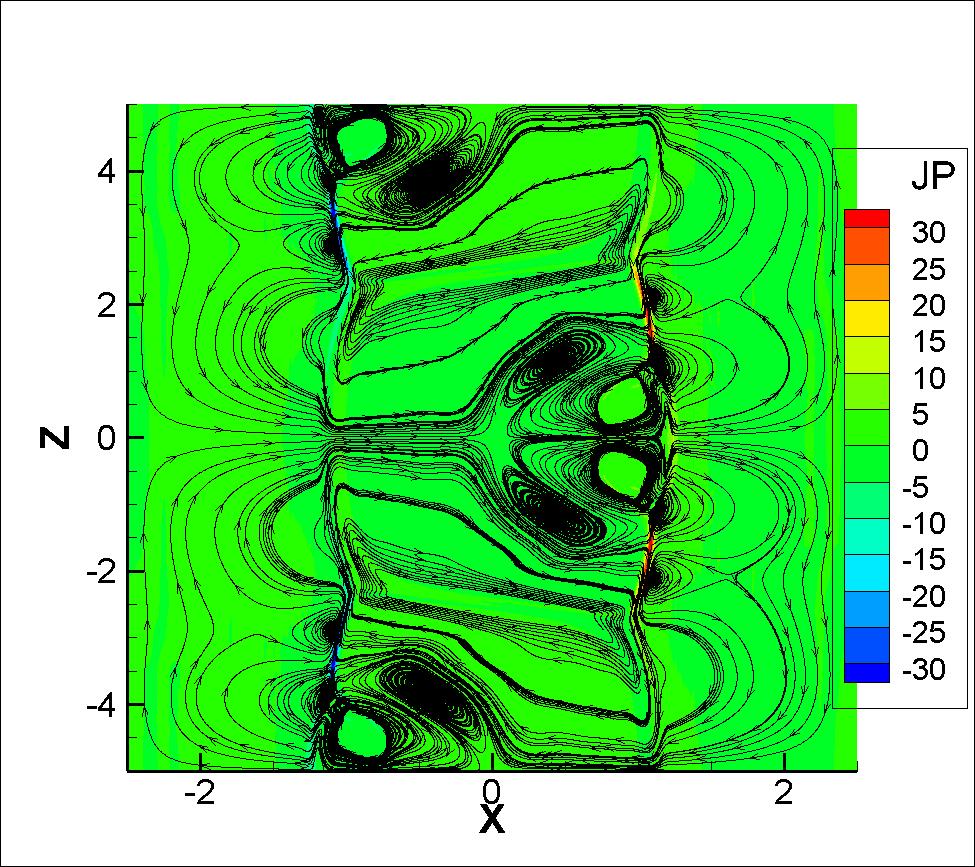}
\put(-175,140){\textbf{(c) t = 100}}
\end{minipage}
\begin{minipage}{0.38\textwidth}
\includegraphics[width=1.0\textwidth]{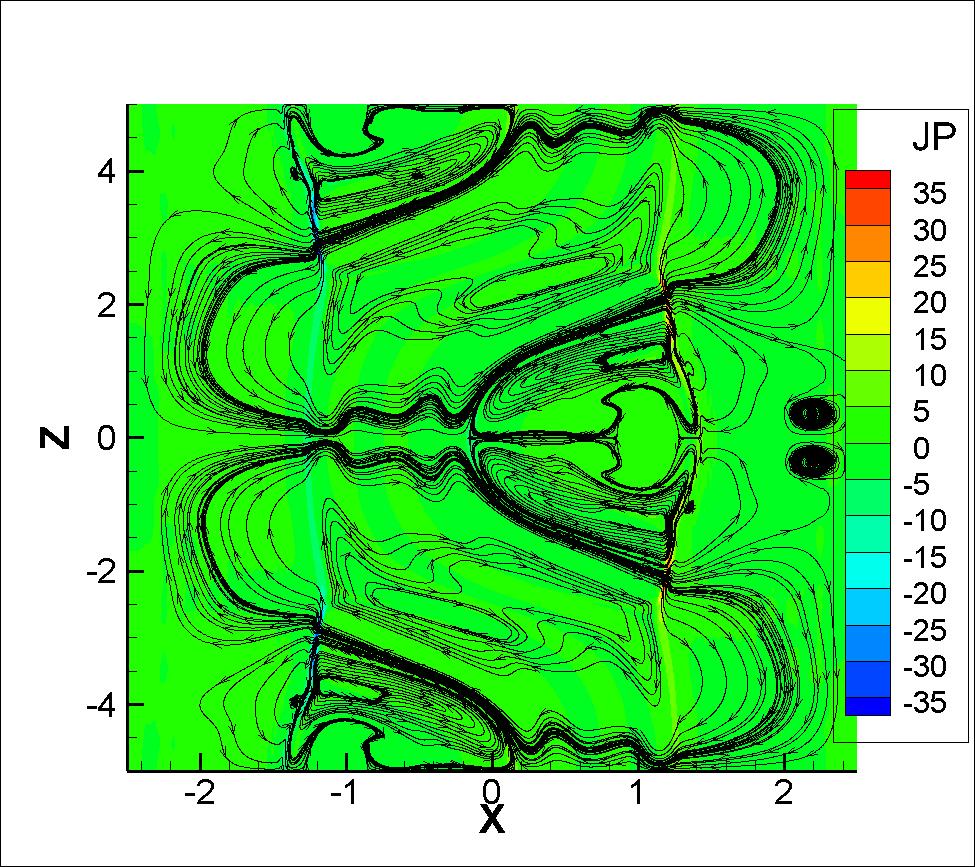}
\put(-175,140){\textbf{(d) t = 115}}
\end{minipage}
\begin{minipage}{0.38\textwidth}
\includegraphics[width=1.0\textwidth]{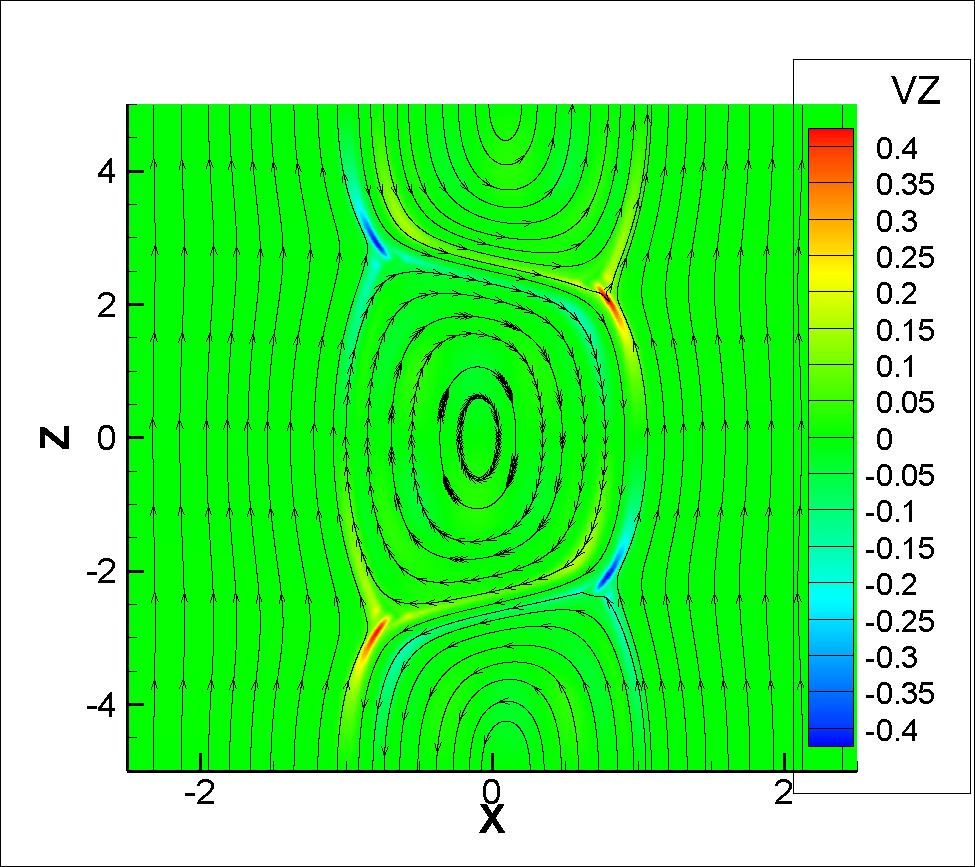}
\put(-175,140){\textbf{(e) t = 85}}
\end{minipage}
\begin{minipage}{0.38\textwidth}
\includegraphics[width=1.0\textwidth]{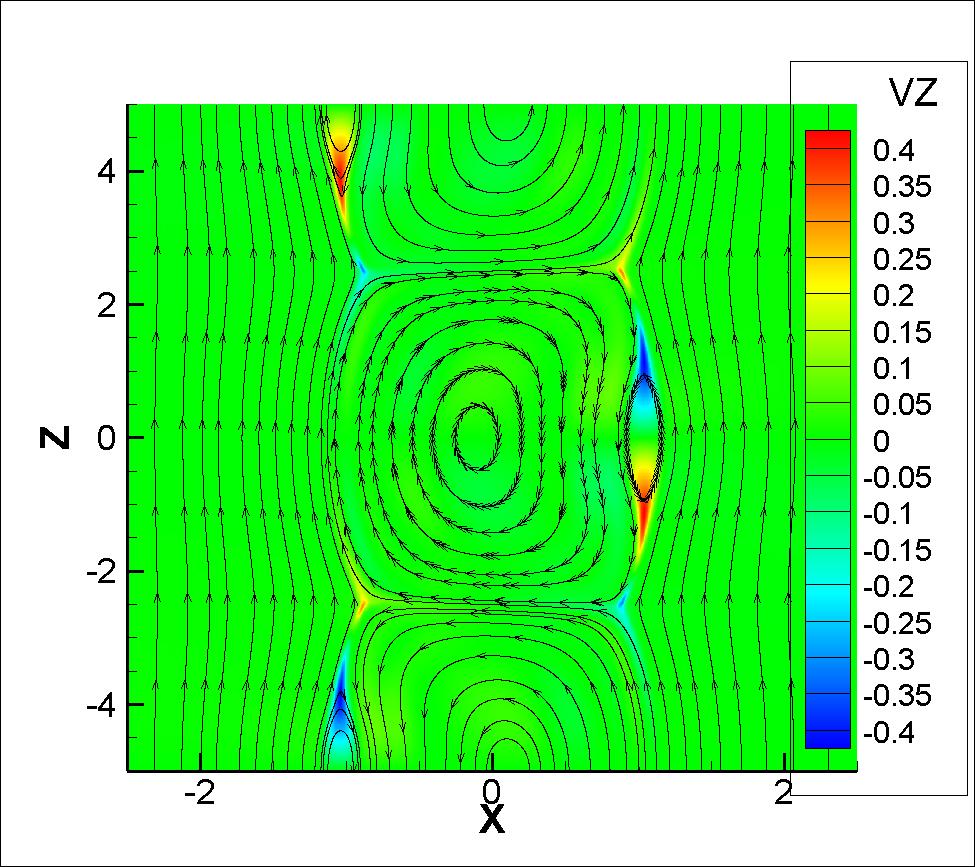}
\put(-175,140){\textbf{(f) t = 95}}
\end{minipage}
\begin{minipage}{0.38\textwidth}
\includegraphics[width=1.0\textwidth]{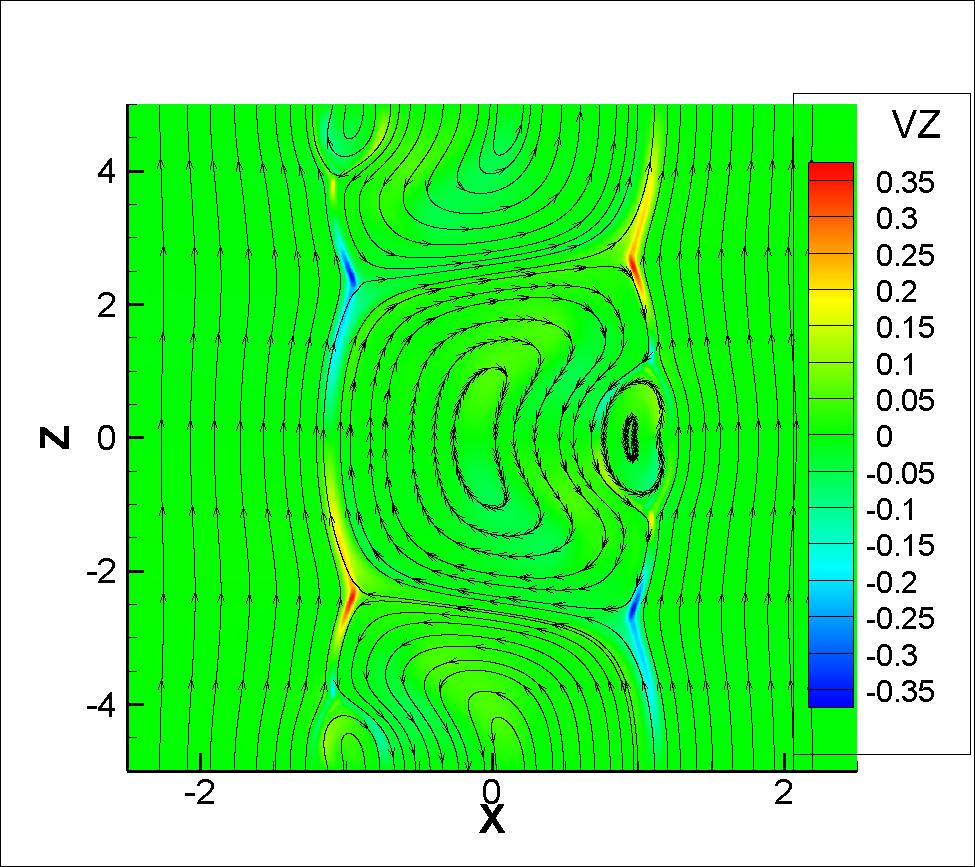}
\put(-175,140){\textbf{(g) t = 100}}
\end{minipage}
\begin{minipage}{0.38\textwidth}
\includegraphics[width=1.0\textwidth]{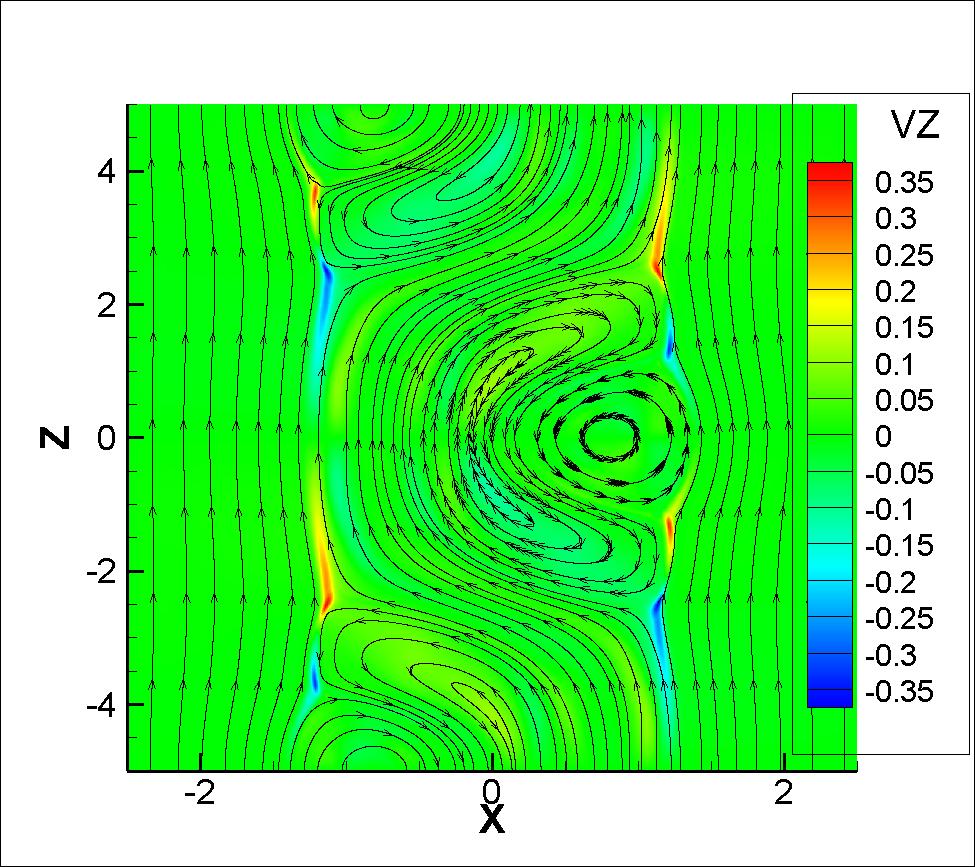}
\put(-175,140){\textbf{(h) t = 115}}
\end{minipage}
\caption{2D contours of the current density in z-direction and flow field stream lines (a-d), 2D contours of the flow field in z-direction and 2D magnetic field lines (e-h), Pr = 1 with ${d = 1.1}$.}
\end{figure}

\begin{figure}[htbp]
\centering
\begin{minipage}{0.38\textwidth}
\includegraphics[width=1.0\textwidth]{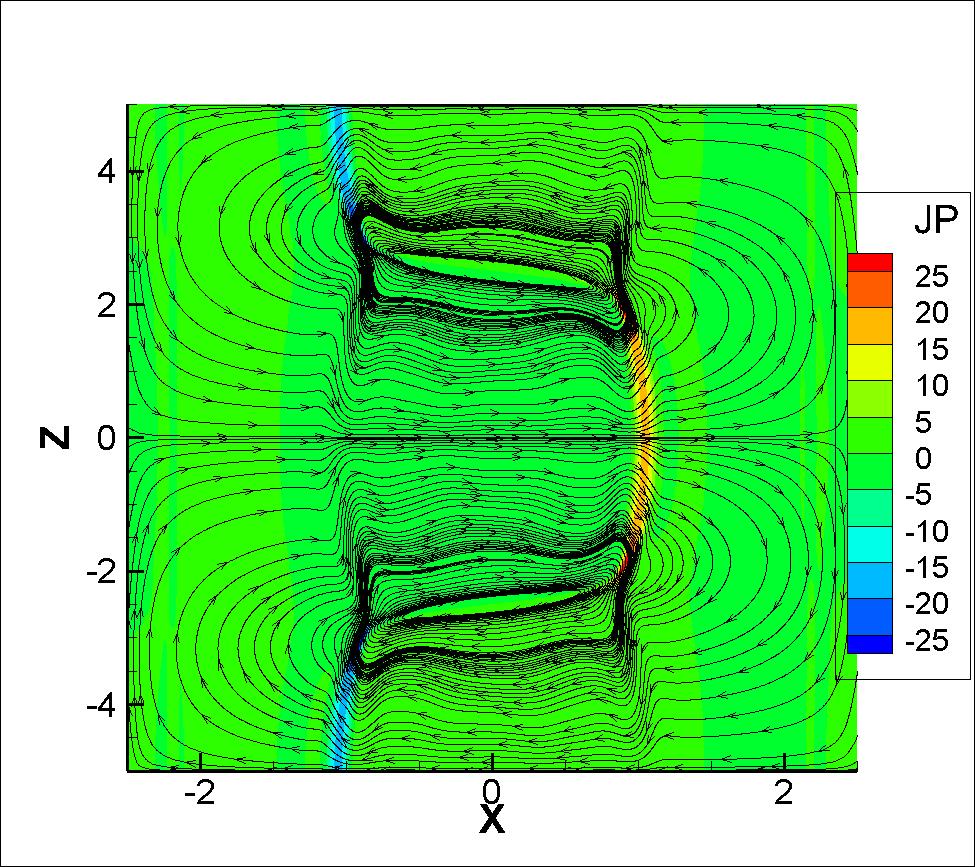}
\put(-175,140){\textbf{(a) t = 116}}
\end{minipage}
\begin{minipage}{0.38\textwidth}
\includegraphics[width=1.0\textwidth]{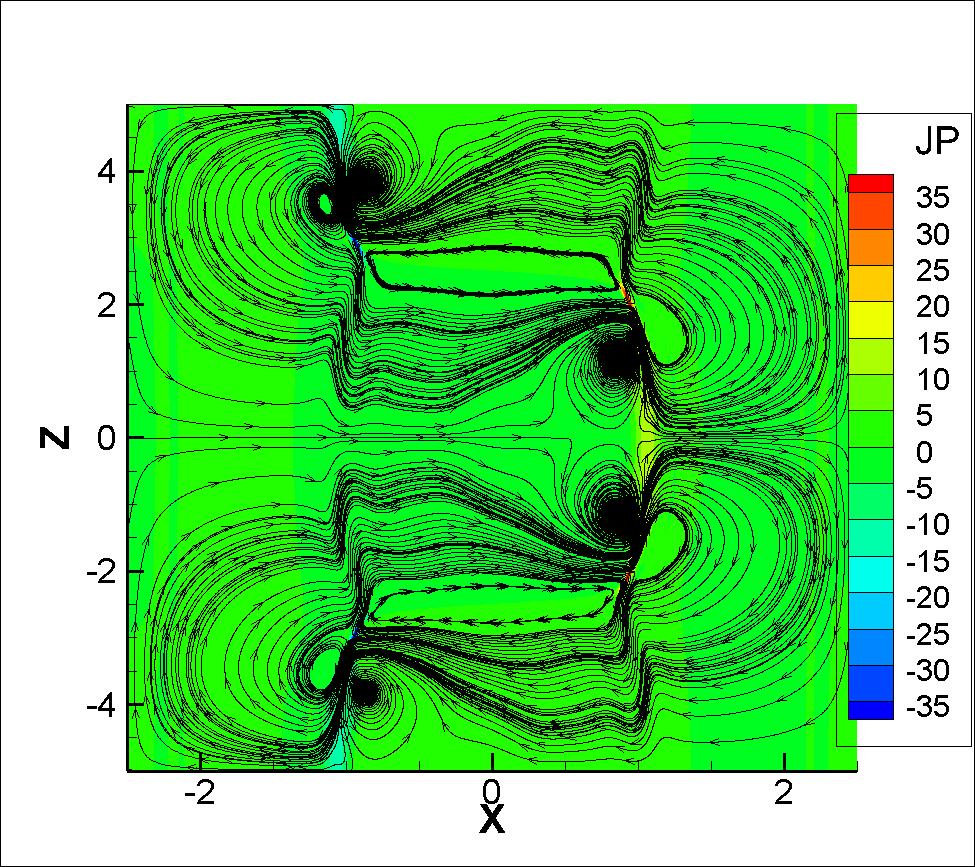}
\put(-175,140){\textbf{(b) t = 120}}
\end{minipage}
\begin{minipage}{0.38\textwidth}
\includegraphics[width=1.0\textwidth]{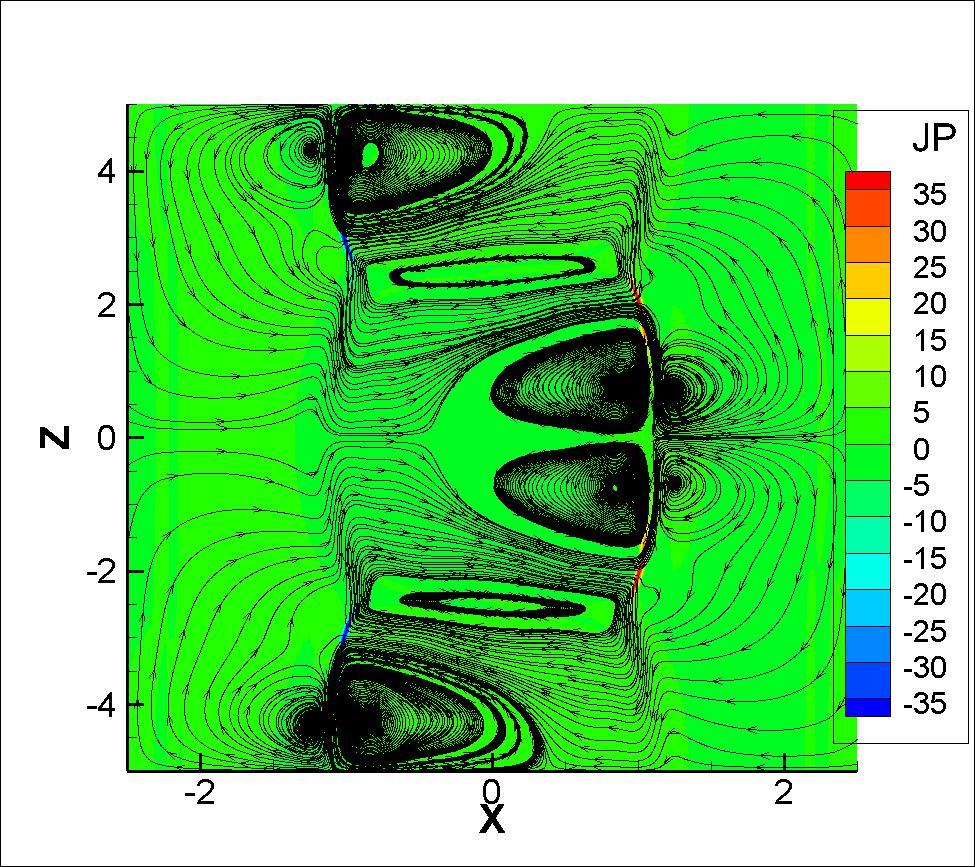}
\put(-175,140){\textbf{(c) t = 124}}
\end{minipage}
\begin{minipage}{0.38\textwidth}
\includegraphics[width=1.0\textwidth]{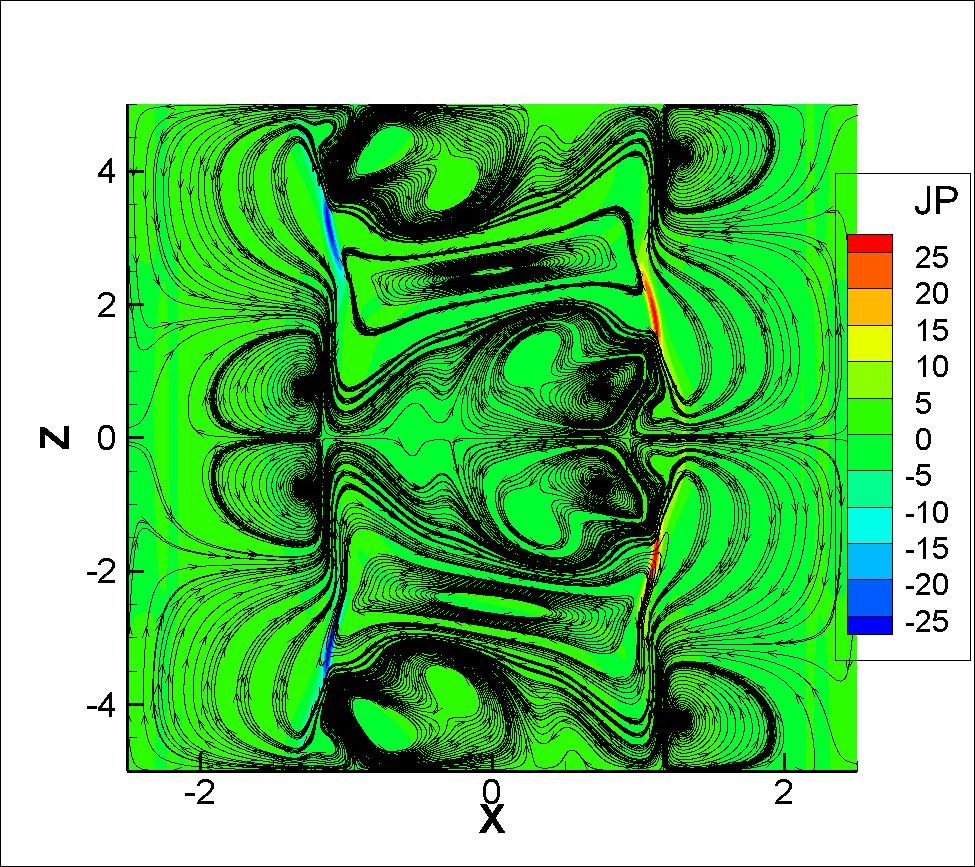}
\put(-175,140){\textbf{(d) t = 132}}
\end{minipage}
\begin{minipage}{0.38\textwidth}
\includegraphics[width=1.0\textwidth]{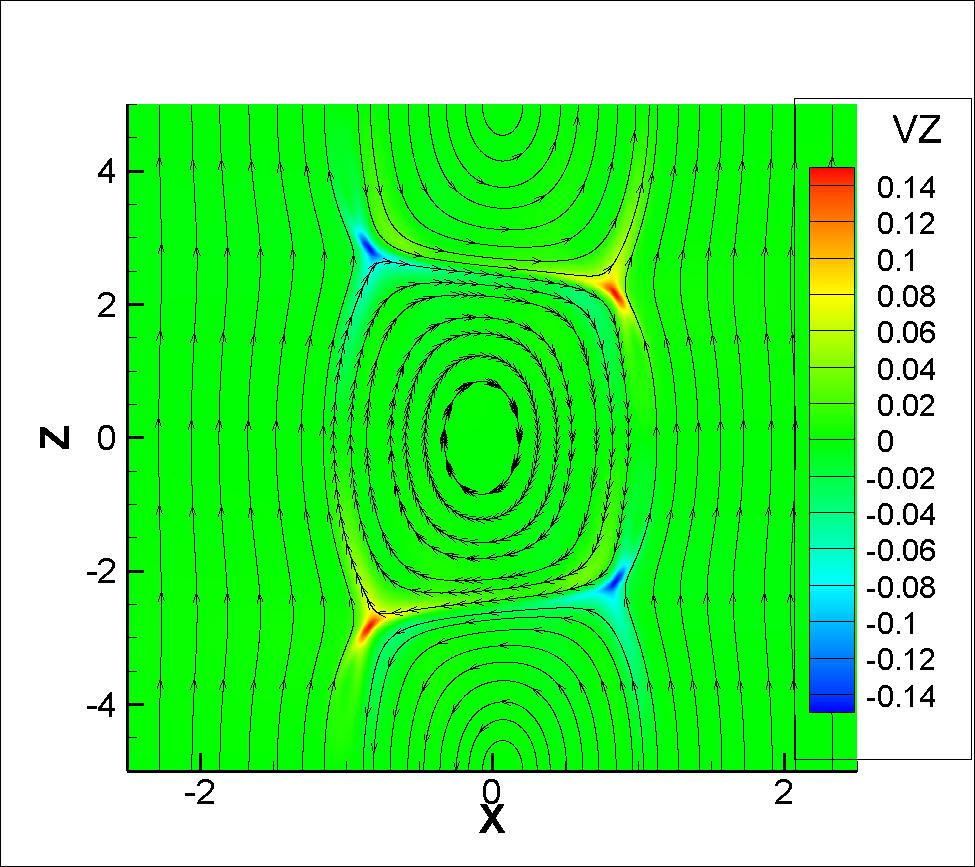}
\put(-175,140){\textbf{(e) t = 116}}
\end{minipage}
\begin{minipage}{0.38\textwidth}
\includegraphics[width=1.0\textwidth]{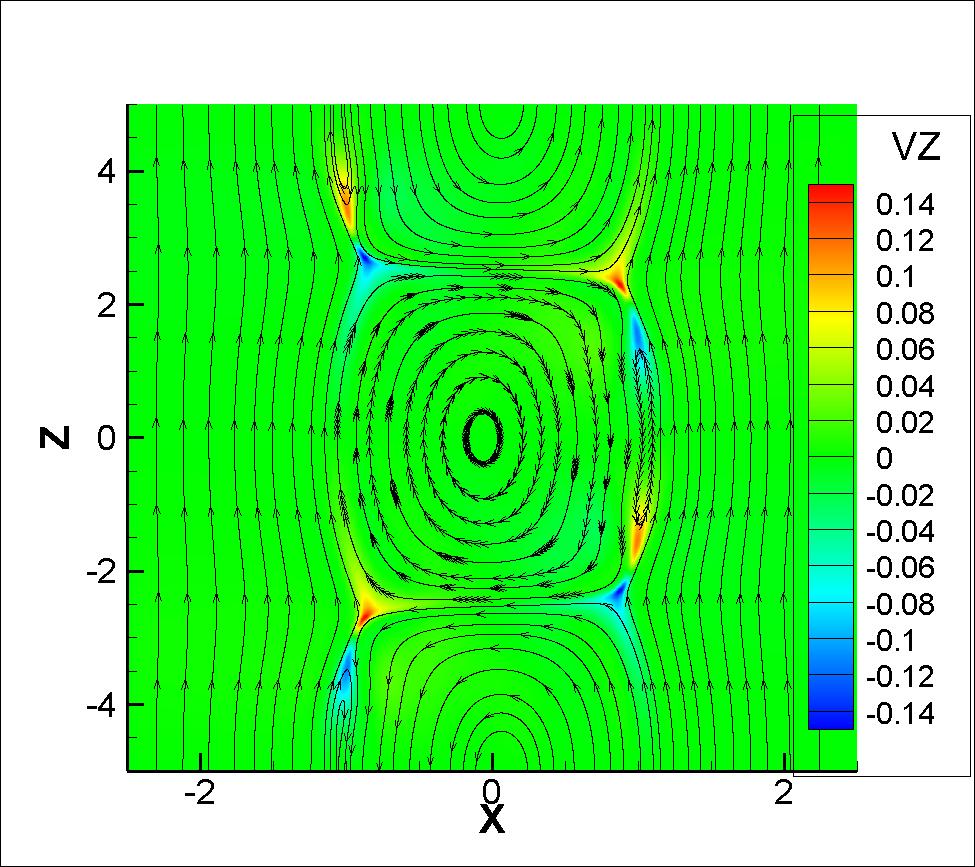}
\put(-175,140){\textbf{(f) t = 120}}
\end{minipage}
\begin{minipage}{0.38\textwidth}
\includegraphics[width=1.0\textwidth]{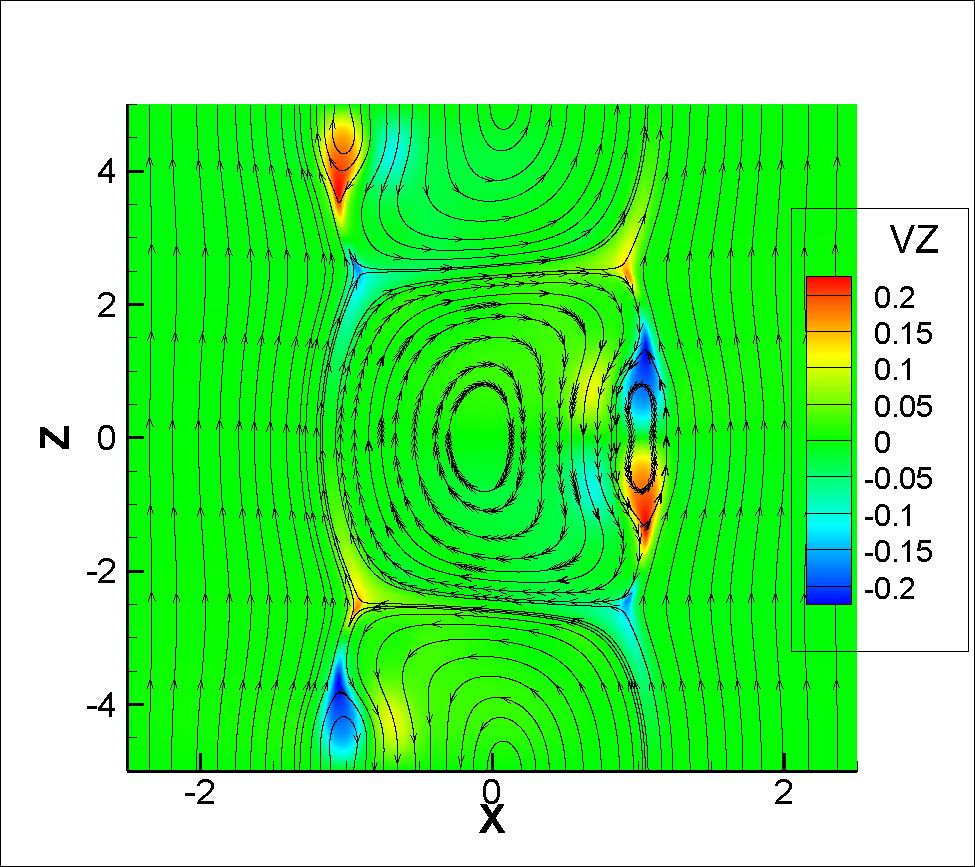}
\put(-175,140){\textbf{(g) t = 124}}
\end{minipage}
\begin{minipage}{0.38\textwidth}
\includegraphics[width=1.0\textwidth]{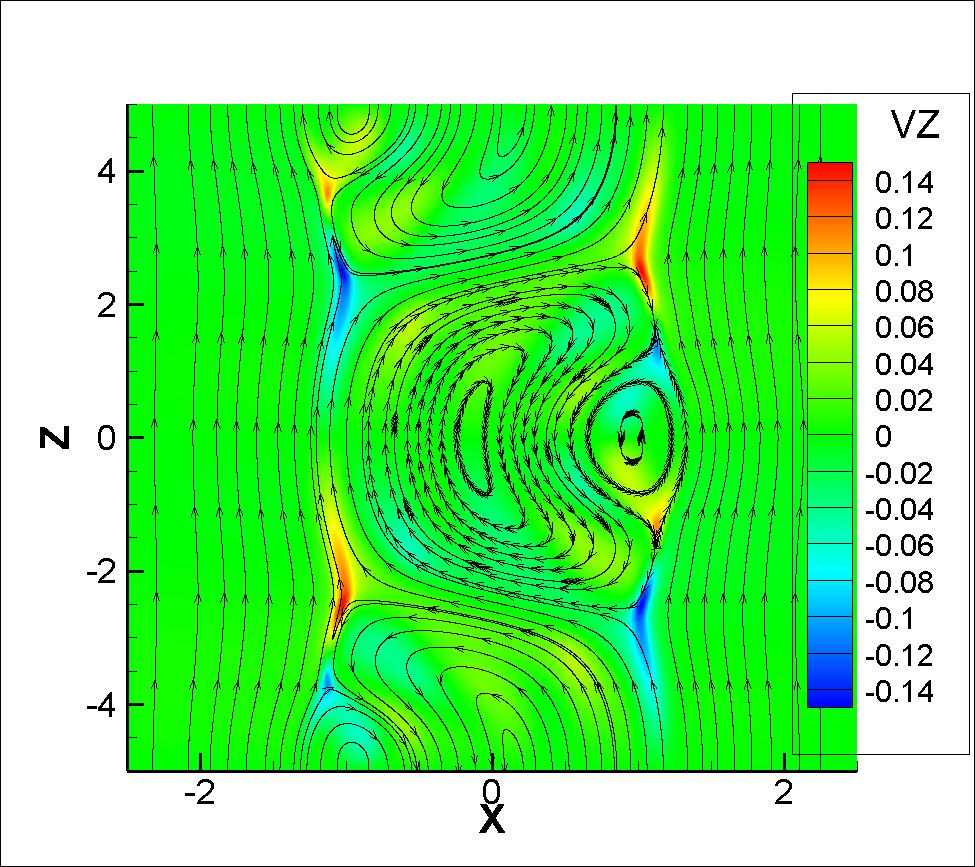}
\put(-175,140){\textbf{(h) t = 132}}
\end{minipage}
\caption{2D contours of the current density in z-direction and flow field stream lines (a-d), 2D contours of the flow field in z-direction and 2D magnetic field lines (e-h), Pr = 10 with ${d = 1.1}$.}
\end{figure}

\begin{figure}[htbp]
\centering
\begin{minipage}{0.38\textwidth}
\includegraphics[width=1.0\textwidth]{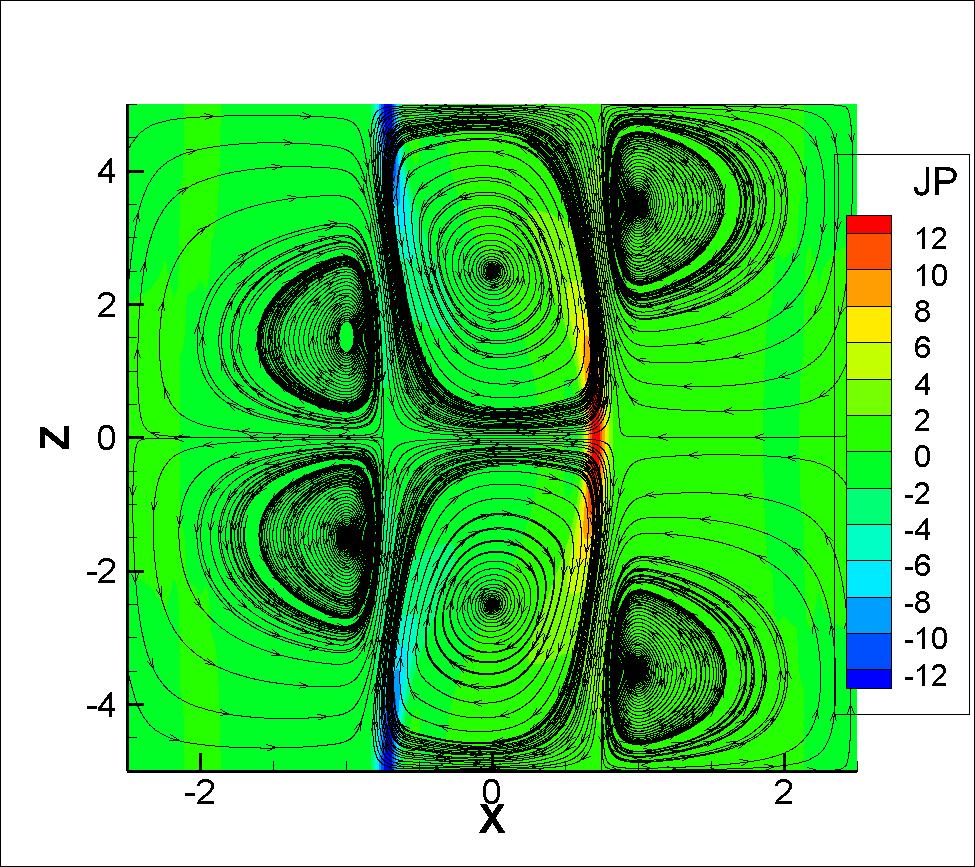}
\put(-175,140){\textbf{(a) t = 160}}
\end{minipage}
\begin{minipage}{0.38\textwidth}
\includegraphics[width=1.0\textwidth]{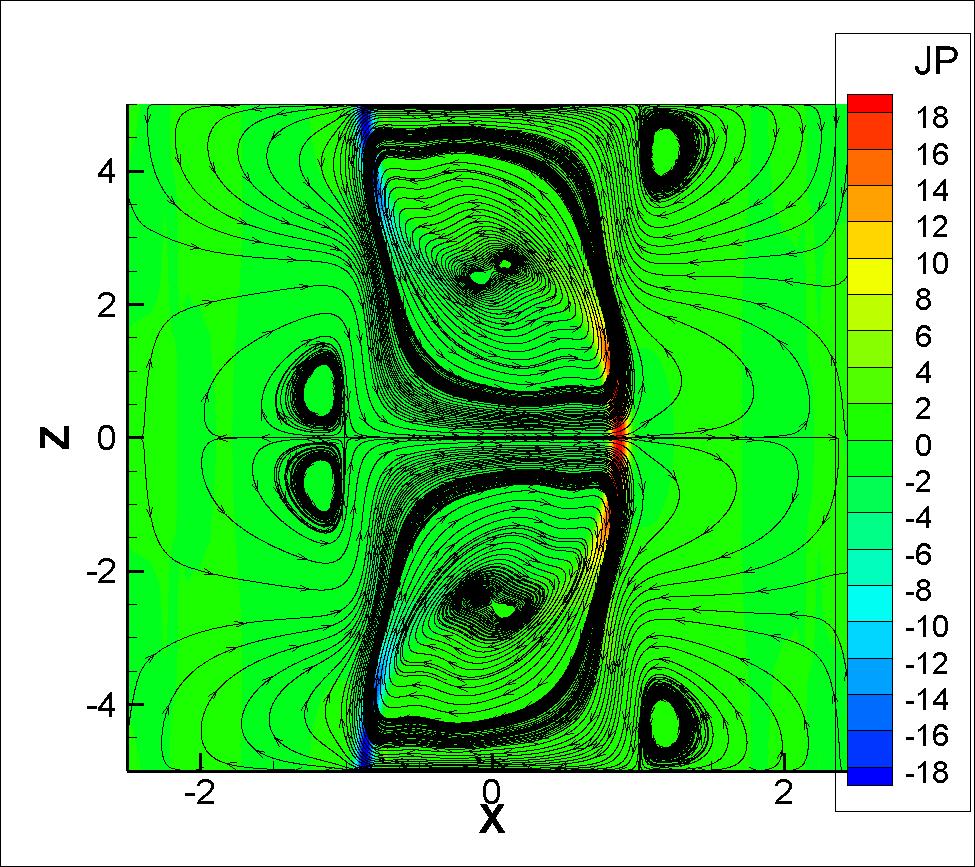}
\put(-175,140){\textbf{(b) t = 185}}
\end{minipage}
\begin{minipage}{0.38\textwidth}
\includegraphics[width=1.0\textwidth]{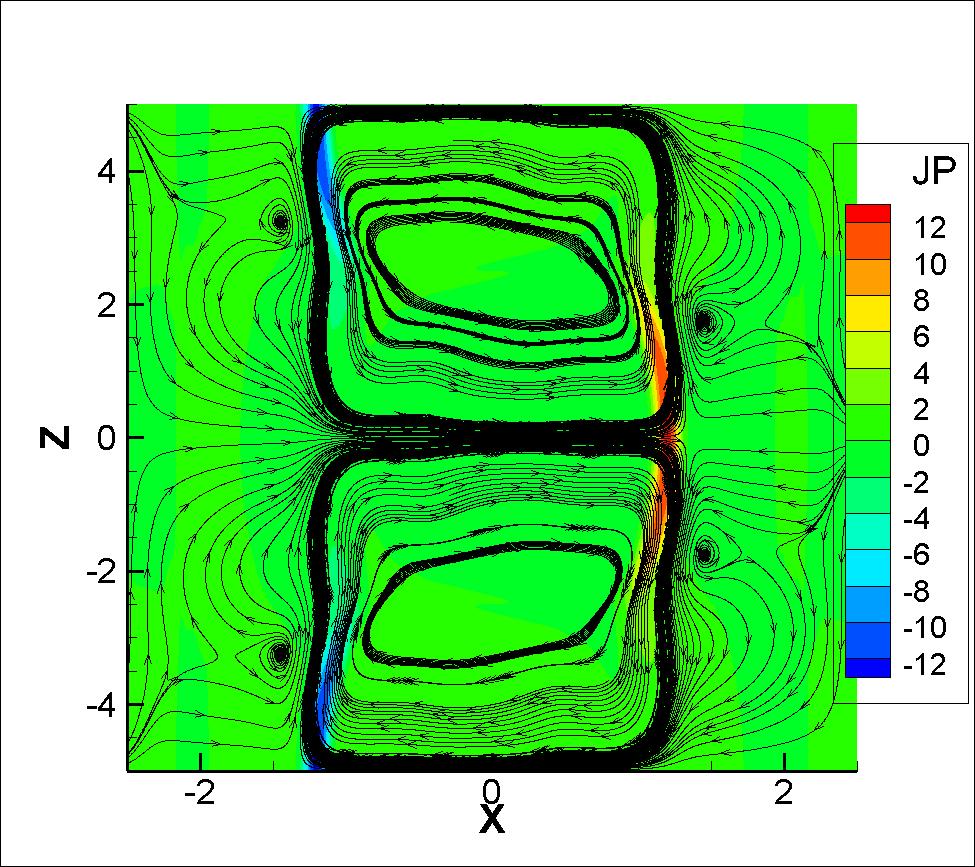}
\put(-175,140){\textbf{(c) t = 250}}
\end{minipage}
\begin{minipage}{0.38\textwidth}
\includegraphics[width=1.0\textwidth]{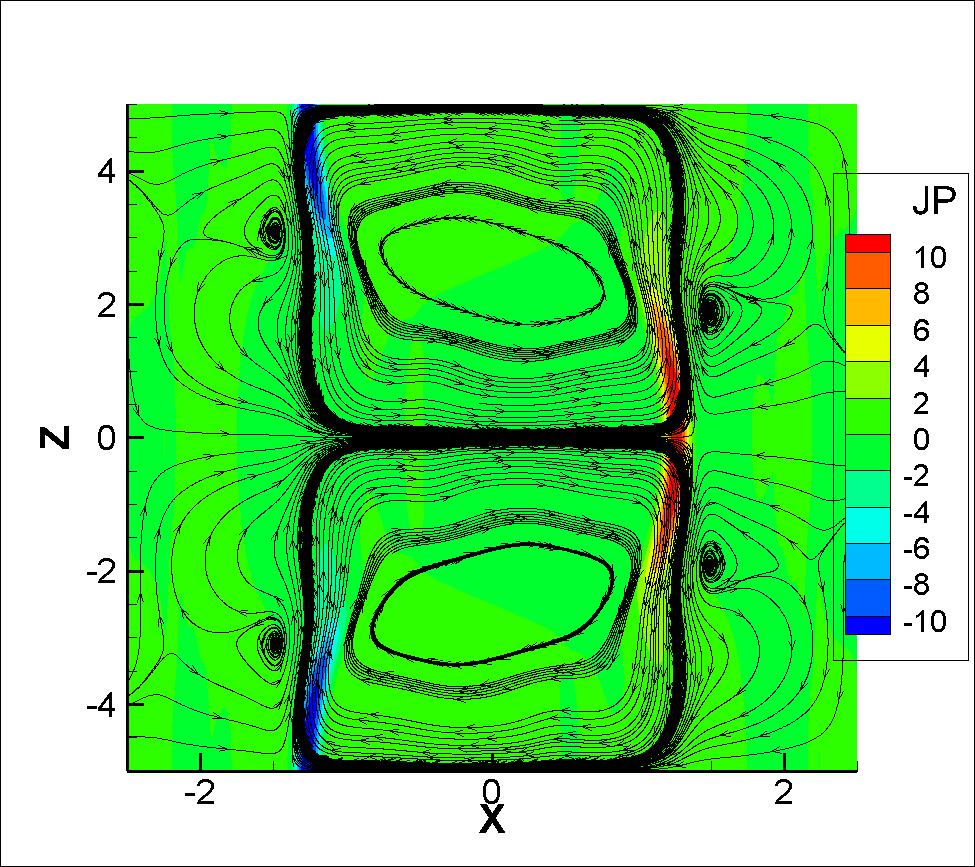}
\put(-175,140){\textbf{(d) t = 270}}
\end{minipage}
\begin{minipage}{0.38\textwidth}
\includegraphics[width=1.0\textwidth]{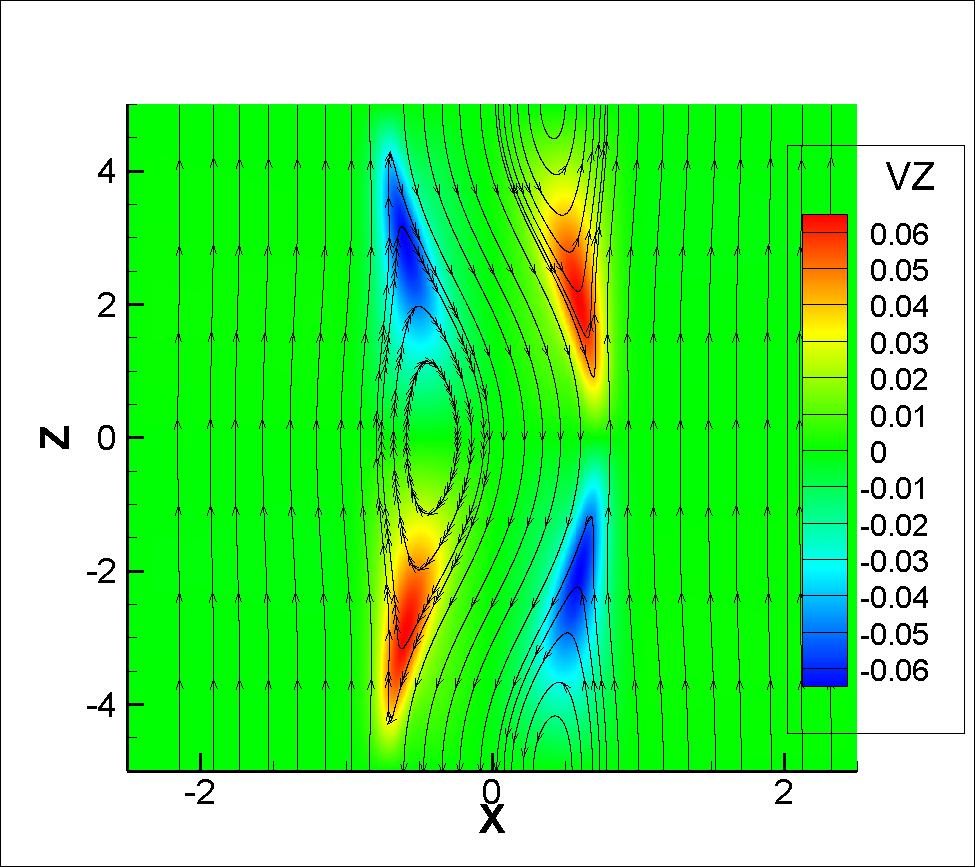}
\put(-175,140){\textbf{(e) t = 160}}
\end{minipage}
\begin{minipage}{0.38\textwidth}
\includegraphics[width=1.0\textwidth]{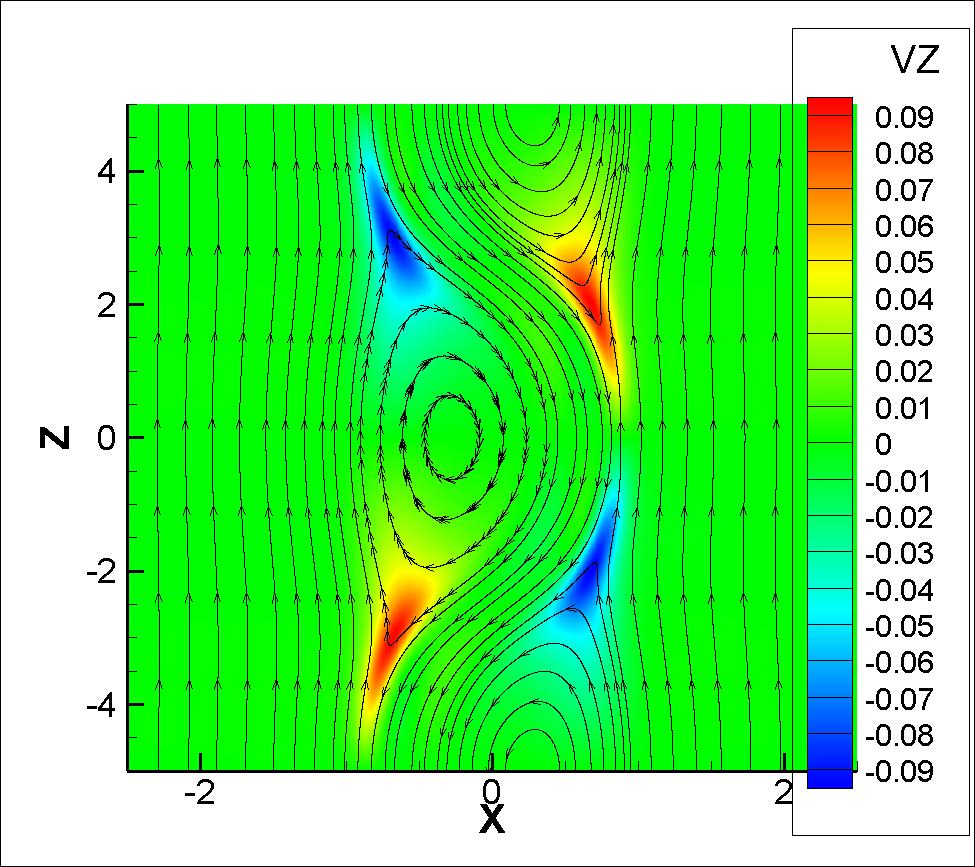}
\put(-175,140){\textbf{(f) t = 185}}
\end{minipage}
\begin{minipage}{0.38\textwidth}
\includegraphics[width=1.0\textwidth]{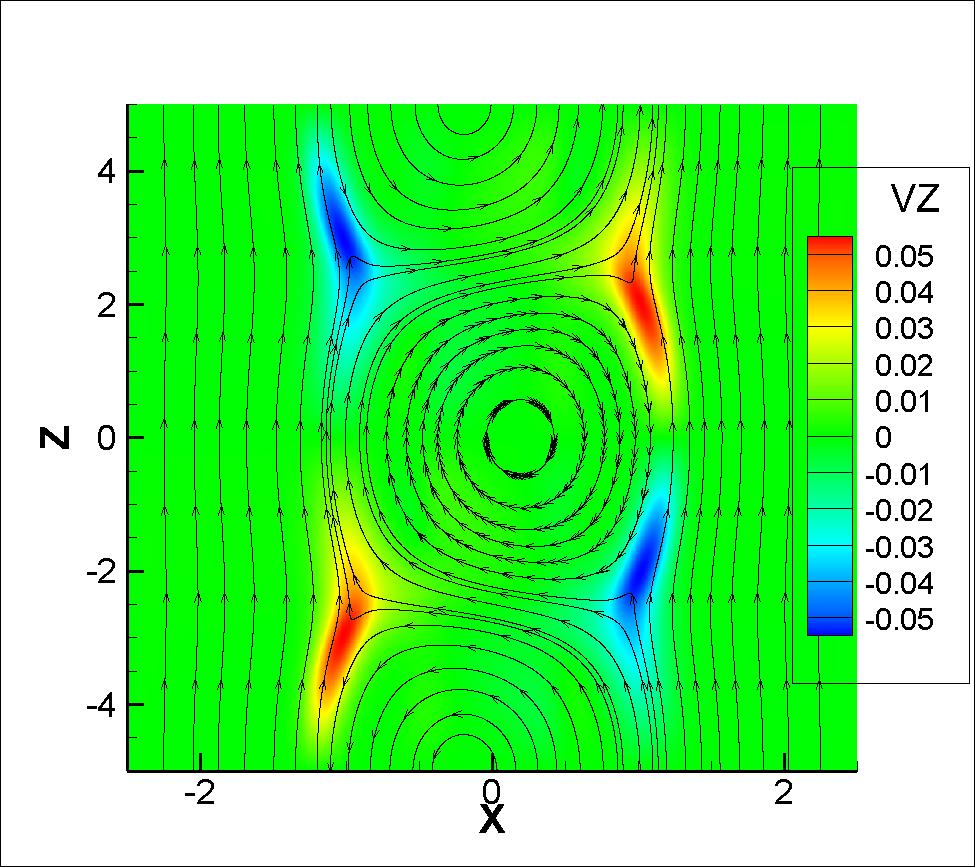}
\put(-175,140){\textbf{(g) t = 250}}
\end{minipage}
\begin{minipage}{0.38\textwidth}
\includegraphics[width=1.0\textwidth]{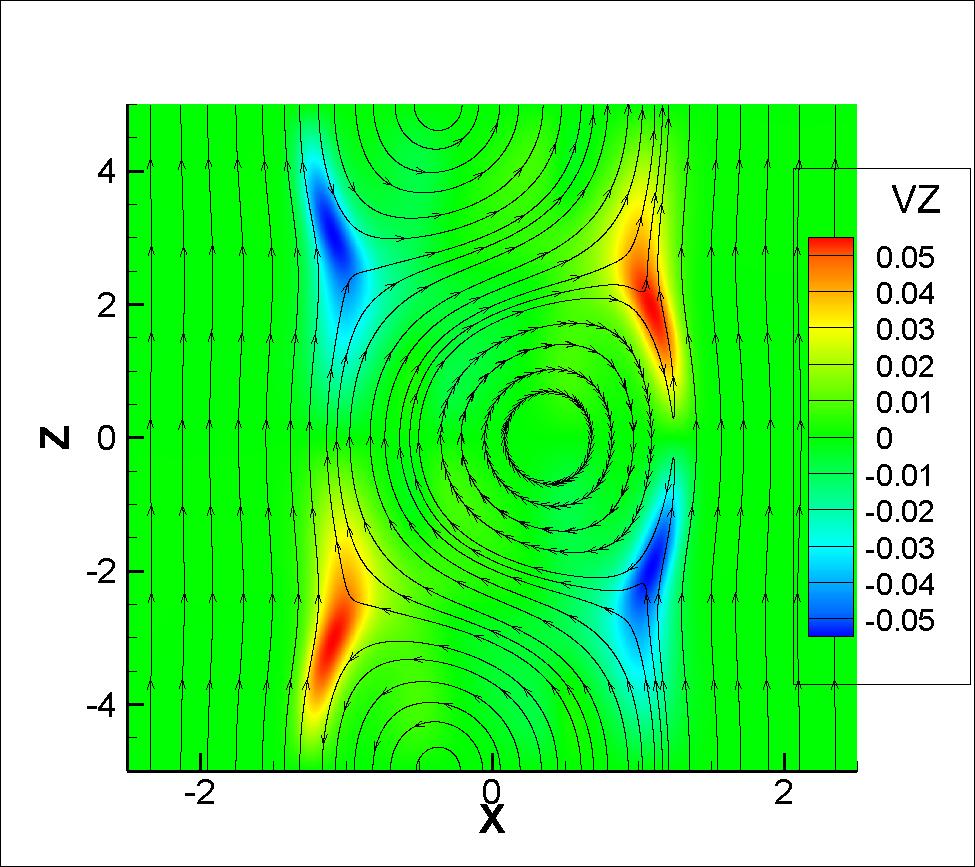}
\put(-175,140){\textbf{(h) t = 270}}
\end{minipage}
\caption{2D contours of the current density in z-direction and flow field stream lines (a-d), 2D contours of the flow field in z-direction and 2D magnetic field lines (e-h), Pr =100 with ${d = 1.1}$.}
\end{figure}

%\bibliography{your-bib-file}
%\bibliography{aipsamp.bib}

\begin{thebibliography}{10}
\expandafter\ifx\csname url\endcsname\relax
  \def\url#1{{\tt #1}}\fi
\expandafter\ifx\csname urlprefix\endcsname\relax\def\urlprefix{URL }\fi
\providecommand{\eprint}[2][]{\url{#2}}

%============================
%======== ref-1 ==========
%\bibitem{hender2007}
%Hender T C \textit{et al} 2007 {\em Nucl. Fusion\/} {\bf 47} S128
\bibitem{bierwage2005fast}
Bierwage A \textit{et al} 2005 \textit{Phys. Plasmas} {\bf 12} 082504
%[1] F. MALARA, P. VELTRI, and V. CARBONE, Phys. Fluids B 4, 3072–3086 (1992).
\bibitem{wang2011interlocking}
Wang X-Q \textit{et al} 2011 \textit {Phys. Plasmas} {\bf 18} 012102
%[2] N. F. LOUREIRO, S. C. COWLEY, W. D. DORLAND, M. G. HAINES, and A. A. SCHEKOCHIHIN, Phys. Rev. Lett. 95, 235003 (2005).
\bibitem{ishii2009plasma}
Ishii Y 2009 \textit {Nucl. Fusion} {\bf 49} 085006
%[3] G. Lapenta, Phys. Rev. Lett. 100, 235001 (2008).
\bibitem{akramov2017non}
Akramov T and Baty H 2017 \textit {Phys. Plasmas} {\bf 24} 082116
%[4] A. Bhattacharjee, Y. Huang, H. Yang, and B. Rogers, Phys. Plasmas 16, 112102 (2009).
\bibitem{dahlburg1995triple}
Dahlburg R B and Karpen J T 1995 \textit {J. Geophys. Res.} {\bf 100} 23489
%[5] Loureiro, N. F., Schekochihin, A. A., & Uzdensky, D. A. 2007, PhPl, 14, 100703
\bibitem{crooker1993multiple}
Crooker N U \textit{et al} 1993 \textit {J. Geophys. Res.} {\bf 98} 9371
%[6] Y. Huang and A. Bhattacharjee, Phys. Plasmas 17, 062104 (2010).
\bibitem{mikic1988dynamical}
Mikic Z \textit{et al} 1988 \textit {Astrophys. J.} {\bf 328} 830
%[7] Huang, Y.-M., & Bhattacharjee, A. 2013, PhPl, 20, 055702
\bibitem{schwartz1988active}
Schwartz S J \textit{et al} 1988 \textit {J. Geophys. Res.} {\bf 93} 11295
%[8] Huang, Y.-M., Bhattacharjee, A., & Sullivan, B. P. 2011, PhPl, 18, 072109
\bibitem{yan1994tearing}
Yan M \textit{et al} 1994 \textit {J. Geophys. Res.} {\bf 99} 8657
%[9] Loureiro, N. F., Schekochihin, A. A., & Uzdensky, D. A. 2013, PhRvE, 87, 013102.
\bibitem{bowers2007spectral}
Bowers K and Li H 2007 \textit {Phys. Rev. Lett.} {\bf 98} 035002
%[10] A. ALI, LI. JIQUAN, and Y. KISHIMOTO, Phys. Plasmas 21, 052312 (2014).
\bibitem{baty2017explosive}
Baty H 2017 \textit {Astrophys. J.} {\bf 837} 74
%[11] Uzdensky, D.A. and Loureiro, N.F., Phys. Rev. Lett. 116, 105003 (2016).
\bibitem{levinton1995improved}
Levinton F M \textit{et al} 1995 \textit {Phys. Rev. Lett.} {\bf 75} 4417
%[12] N. F. Loureiro, R. Samtaney, A. A. Schekochihin, and D. A. Uzdensky. Phys. Plasmas 19, 042303 (2012).
\bibitem{ishii2002structure}
Ishii Y \textit{et al} 2002 \textit {Phys. Rev. Lett.} {\bf 89} 205002
%[13] DOBROTT, D., PRAGER, S. C. & TAYLOR, J. B. 1977 Phys. Fluids 20, 1850–1854.

\bibitem{wang2007fast}
Wang Z X \textit{et al} 2007 \textit {Phys. Rev. Lett.} {\bf 99} 185004
%[14] BULANOV, S. V., SYROVATSKIˇI, S. I. & SAKAI, J. 1978 J. Expl Theor. Phys. 28, 177.
\bibitem{janvier2011structure}
Janvier M \textit{et al} 2011 \textit {Phys. Rev. Lett.} {\bf 107} 195001
%[15] BULANOV, S. V., SAKAI, J. & SYROVATSKIˇI , S. I. 1979 Sov. J. Plasma Phys. 5, 157–173.
\bibitem{priest1985magnetohydrodynamics}
Priest E R 1985 \textit {Rep. on Progress in Phy.} {\bf 48} 955
\bibitem{chang1996off}
Chang Z \textit{et al} 1996 \textit {Phys. Rev. Lett.} {\bf 77} 3553
%[16] BISKAMP, D. 1986 Phys. Fluids 29, 1520–1531.
\bibitem{ishii2003long}
Ishii Y \textit{et al} 2003 \textit {Nucl. Fusion} {\bf 43} 539
\bibitem{wang2008shear}
Wang Z X \textit{et al} 2008 \textit {Phys. Plasmas} {\bf 15} 082109
\bibitem{pritchett1980linear}
Pritchett P L  \textit{et al} 1980 \textit {Phys. Fluids} {\bf 23} 1368
\bibitem{yu1996nonlinear}
Yu Q 1996 \textit {Phys. Plasmas} {\bf 3} 2898
%[27] L. Comisso , M. Lingam, Y.-M. Huang, and A. Bhattacharjee The Astrophysical Journal, 850:142 (16pp), 2017 December 1.
%\bibitem{Goncalves2018}
%Gonçalves B \textit{et al} 2018 \textit {arXiv e-printsArXiv} {\bf 1804} 07123
%[28] B. Gonçalves, I. Henriques, C. Hidalgo, C. Silva, H. Figueiredo, V. Naulin, A. H. Nielsen, J.T. Mendonça, arXiv e-printsArXiv:1804.07123 (2018).
\bibitem{persson1994nonlinear}
Persson M and Dewar R L 1994 \textit {Phys. Plasmas} {\bf 1} 1256
\bibitem{furth1973tearing}
Furth H P \textit{et al} 1973 \textit {Phys. Fluids} {\bf 16} 1054
%[24] TENERANI, A., RAPPAZZO, A. F., VELLI, M. & PUCCI, F. 2015a Astrophys. J. 801, 145, 1–7.

%[26] Comisso, L., & Grasso, D. 2016, PhPl, 23, 032111.
%[29] R. C. Sovinec, A. H. Glasser, T. H. Gianakon, D. C. Barnes, R. A. Nebel, S. E. Kruger, D. D. Schnack, S. J. Plimpton, A. Tarditi, and M. S. Chu, J. Comput. Phys. 195, 355 (2004).
\bibitem{dong2003double}
Dong J Q \textit{et al} 2003 \textit {Phys. Plasmas} {\bf 10} 3151
%[30] E. G. Harris, Nuovo Cimento 23, 115–121 (1962).
\bibitem{held1999magnetohydrodynamic}
Held E D \textit{et al} 1999 \textit {Phys. Plasmas} {\bf 6} 837
\bibitem{ma2017effect}
Ma J \textit{et al} 2017 \textit{Nucl. Fusion} {\bf 57} 126004
\bibitem{zhang2011nonlinear}
Zhang C L and Ma Z W 2011 \textit {Phys. Plasmas} {\bf 18} 052303
\bibitem{nemati2018unstable}
Nemati M J \textit{et al} 2018 \textit {Phys. Plasmas} {\bf 25} 072119

\bibitem{shen1998properties}
Shen C \textit{et al} 1998 \textit{Phys. Lett. A} {\bf 249} 87
\bibitem{shen1998magnetic}
Shen C and Li Z X 1998 \textit{Plasma Phys. Control. Fusion} {\bf 40} 1
\bibitem{nemati2016formation}
Nemati M J \textit{et al} 2016 \textit {Astrophys. J.} {\bf 821} 128
\bibitem{nemati2017dynamics}
Nemati M J \textit{et al} 2017 \textit {Astrophys. J.} {\bf 835} 191
\bibitem{Guo2017}
Guo W \textit{et al} 2017 \textit {Phys. Plasmas} {\bf 24} 032115
\bibitem{konovalov2002transport}
Konovalov S V  \textit{et al} 2002 \textit {Phys. Plasmas} {\bf 9} 4596
\bibitem{kaw1979tearing}
Kaw P K \textit{et al} 1979 \textit{Phys. Rev. Lett.} {\bf43} 1398
\bibitem{ahmad2021viscous}
Ahmad N \textit{et al} 2022 \textit{Plasma Sci. Technol.} {\bf24} 015103
\bibitem{ALI2015}
Ali A \textit{et al} 2015 \textit {Phys. Plasmas} {\bf 22} 042102
\bibitem{ALI2019}
Ali A and Zhu P 2019 \textit {Phys. Plasmas} {\bf 26} 052518
\bibitem{aydemir1990magnetohydrodynamic}
Aydemir A Y 1990 \textit {Phys. Fluids B} {\bf 2} 2135
\bibitem{sovinec2004nonlinear}
Sovinec R C \textit{et al} 2004 \textit{J. Comput. Phys.} {\bf195}, 355.

\bibitem{He2015}
He Z \textit{et al} 2015 \textit{Phys. Scr.} {\bf90}, 125603.
\bibitem{Mao2021}
Mao A \textit{et al} 2021 \textit{Plasma Sci. Technol.} {\bf23}, 035103.

\bibitem{coppi1976resistive}
Coppi B \textit{et al} 1976 \textit {Sov. J. Plasma Phys.} {\bf 2} 533
\bibitem{Shu2013}

Wei L  \textit{et al} 2012 \textit {J. Plasma Phys.} {\bf 78} 663

\bibitem{ofman1992double}
Ofman L 1992 \textit {Phys. Fluids B} {\bf 4} 2751
\bibitem{ali2014abrupt}
Ali A \textit{et al} 2014 \textit {Phys. Plasmas} {\bf 21} 052312
%[25] H. Betar, D. Del Sarto, M. Ottaviani, and A. Ghizzo Phys. Plasmas 27, 102106 (2020).

%[22] GRASSO, D., HASTIE, R. J., PORCELLI, F. & TEBALDI, C. 2008 Phys. Plasmas 15, 072113.
%\bibitem{nemati2015plasmoid}
%Nemati M J \textit{et al} 2015 \textit {Phys. Plasmas} {\bf 22} 012106
%[23] MILITELLO, F., BORGOGNO, D., GRASSO, D., MARCHETTO, C. & OTTAVIANI, M. 2011 Phys. Plasmas 18, 112108.

%[31] A. Tenerani, M. Velli, F. Pucci, S. Landi and A. F. Rappazzo, J. Plasma Phys. 5, 82 (2016).


%[32] B. Coppi, E. Galvao, R. Pellat, M. N. Rosenbluth, and P. H. Rutherford, Sov. J. Plasma Phys. 2, 533 (1976).

%\bibitem{Steinolfson1984}
 %Steinolfson R S and Van Hoven G 1984 \textit {Phys. Fluids} {\bf 27} 1207

\end{thebibliography}

\end{document}